\documentclass{aa}
\usepackage[varg]{txfonts}
\usepackage{wasysym}
\usepackage{graphicx}
\usepackage{gensymb}
\usepackage{subcaption}
\usepackage{arydshln}
\usepackage{rotating}

\begin{document}

\title{Properties of stream interaction regions at Earth and Mars during the declining phase of SC 24}
\author{Paul Geyer\inst{\ref{inst1}}
\and Manuela Temmer\inst{\ref{inst1}}
\and Jingnan Guo\inst{\ref{inst2}, \ref{inst3}}
\and Stephan G. Heinemann \inst{\ref{inst1},\ref{inst4}}}

\institute{University of Graz, Institute of Physics, Austria \label{inst1}
\and School of Earth and Space Sciences, University of Science and Technology of China, Hefei, PR China
\label{inst2}
\and CAS Center for Excellence in Comparative Planetology, Hefei, PR China
\label{inst3}
\and Max-Planck-Institut für Sonnensystemforschung, Justus-von-Liebig-Weg 3, 37077 Göttingen, Germany
\label{inst4}}


\date{\today}

\abstract{}{We inspect the evolution of stream interaction regions (SIRs) from Earth to Mars, covering the distance range 1--1.5 AU, over the declining phase of solar cycle 24 (2014--2018). So far, studies only analyzed SIRs measured at Earth and Mars at different times. We compare existing catalogs for both heliospheric distances and arrive at a clean dataset for the identical time range. This allows a well-sampled statistical analysis and for the opposition phases of the planets an in-depth analysis of SIRs as they evolve with distance.}{We use in-situ solar wind data from OMNI and the Mars Atmosphere and Volatile EvolutioN (MAVEN) spacecraft as well as remote sensing data from Solar Dynamics Observatory (SDO). A superposed epoch analysis is performed for bulk speed, proton density, temperature, magnetic field magnitude and total perpendicular pressure. Additionally, a study of events during the two opposition phases of Earth and Mars in the years 2016 and 2018 is conducted. SIR related coronal holes with their area as well as their latitudinal and longitudinal extent are extracted and correlated to the maximum bulk speed and duration of the corresponding high speed solar wind streams following the stream interaction regions.}{We find that while the entire solar wind high speed stream shows no expansion as it evolves from Earth to Mars, the crest of the high speed stream profile broadens by about 17\%, and the magnetic field and total pressure by about 45\% around the stream interface. The difference between the maximum and minimum values in the normalized superposed profiles increases slightly or stagnates from 1--1.5 AU for all parameters, except for the temperature. A sharp drop at zero epoch time is observed in the superposed profiles for the magnetic field strength at both heliospheric distances. The two opposition phases reveal similar correlations of in-situ data with coronal hole parameters for both planets. Maximum solar wind speed has a stronger dependence on the latitudinal extent of the respective coronal hole than on its longitudinal extent. We arrive at an occurrence rate of fast forward shocks three times as high at Mars than at Earth.}{}


\maketitle

\section{Introduction} \label{intro}
Solar wind (SW) high speed streams (HSS) emanate from coronal holes (CHs) on the Sun and form with the ambient slow solar wind so-called stream interaction regions (SIRs). As CHs are rather long-lived structures, SIRs can be observed over several rotations and are then called co-rotating interaction regions (CIRs). SIR/CIR structures are detected at various locations in the inner heliosphere by in-situ instruments, showing typical characteristics. As the fast SW (with speeds as high as $800\  \mathrm{km s^{-1}}$) compresses the slow portion (about $400\ \mathrm{km s^{-1}}$), density and interplanetary magnetic field (IMF) magnitude are rapidly enhanced \citep[see e.g.,][]{Schwenn06}. The density finds its maximum just before the total perpendicular pressure has its peak (slow SW changes to fast SW regime) which is commonly referred to as the stream interface (SI) \citep[e.g.,][]{Jian06}. The fast SW propagates freely with its characteristic high speed in the rarefaction region following the SIR. CIRs and SIRs may develop forward or reverse shocks (or both) as they expand through the heliosphere \citep[e.g.,][]{Jian06,Huang19}.

The formation of shocks can be explained using simple 1-D modeling \citep[e.g][]{Hundhausen73}: as the crest of the pressure wave moves faster than the front decelerated by the slow SW ahead, the wave steepens. Eventually, this steepening evolves into a sharp, discrete boundary at the front and rear of the wave, thus forming shocks. Hence, the further out in the heliosphere the higher the occurrence rate of both kinds of shocks, due to further steepening as observed when comparing SIRs at 0.3 AU and 1 AU \citep[see also][]{richter86}. More sophisticated models taking the 3-D evolution of SIRs into account \citep[see e.g.,][]{Gosling99} conclude different occurrence rates for shocks depending on the heliographic latitude. Forward shocks tend to weaken and eventually disappear for latitudes $\gtrapprox 30 \degree$ while the occurrence rate of reverse shocks is still significant for these latitudes. Additionally to this latitudinal dependency, interactions between front and reverse shocks are expected to take place from 10 AU on. A very recent review by \cite{richardson18} gives an overview of various studies of SIRs observed at different radial distances from the Sun.

While the properties of SIRs/CIRs are well documented at 1 AU, SW plasma and magnetic field measurements are performed at very scarcely distributed locations in the heliosphere, especially beyond 1 AU. The Mars Atmosphere and Volatile EvolutioN \citep[MAVEN,][]{Jakosky15} spacecraft launched in November 2013 has been measuring SW plasma and magnetic field at the orbit of Mars to expand our knowledge to the outermost of the inner planets. Among other instruments used for measuring atmospheric particles, MAVEN is carrying three scientific instruments dedicated to SW measurements: the Solar Wind Electron Analyzer \citep[SWEA,][]{Mitchell16}, the Solar Wind Ion Analyzer \citep[SWIA,][]{Halekas15} as well as the Magnetometer \citep[MAG,][]{Connerney15}. After reaching Mars in September 2014, MAVEN was inserted into an elliptical orbit with a period of 4.5 h \citep{Jakosky15}. Data gaps in the observations occur when the spacecraft is in safe mode and during scheduled observation campaigns, which require off-pointing. Besides, MAVEN often traverses through the bow shock of Mars especially when the spacecraft crosses the night-side of the planet. To select the data measured during the solar wind intervals, \citet{Halekas17} developed an algorithm based on the SWIA aboard moments data and MAG measurements to extract solar wind quantities over the upstream segment of each MAVEN orbit. 

\citet{Lee17} and \citet{Huang19} studied SW events (i.e., SIRs/CIRs and coronal mass ejections) at Mars using MAVEN data, also comparing individual events to those measured at Earth. However, the statistical comparison of a larger sample of events measured both at Earth and Mars is still missing. Here, we study for a consistent set of CIRs detected at Earth and Mars over the time span November 2014 -- November 2018 their expansion behavior and physical characteristics over the heliospheric distance range 1 to 1.5~AU. For a subset of aligned events (which occurred during a limited time span in our sample) we investigate the spatial evolution of SIRs/CIRs and HSSs when Earth and Mars were closely aligned longitudinally and compare the results to their solar sources, that is, CHs.

\section{Data and methods} \label{data}

\begin{figure}
\centering
\resizebox{\hsize}{!}{\includegraphics{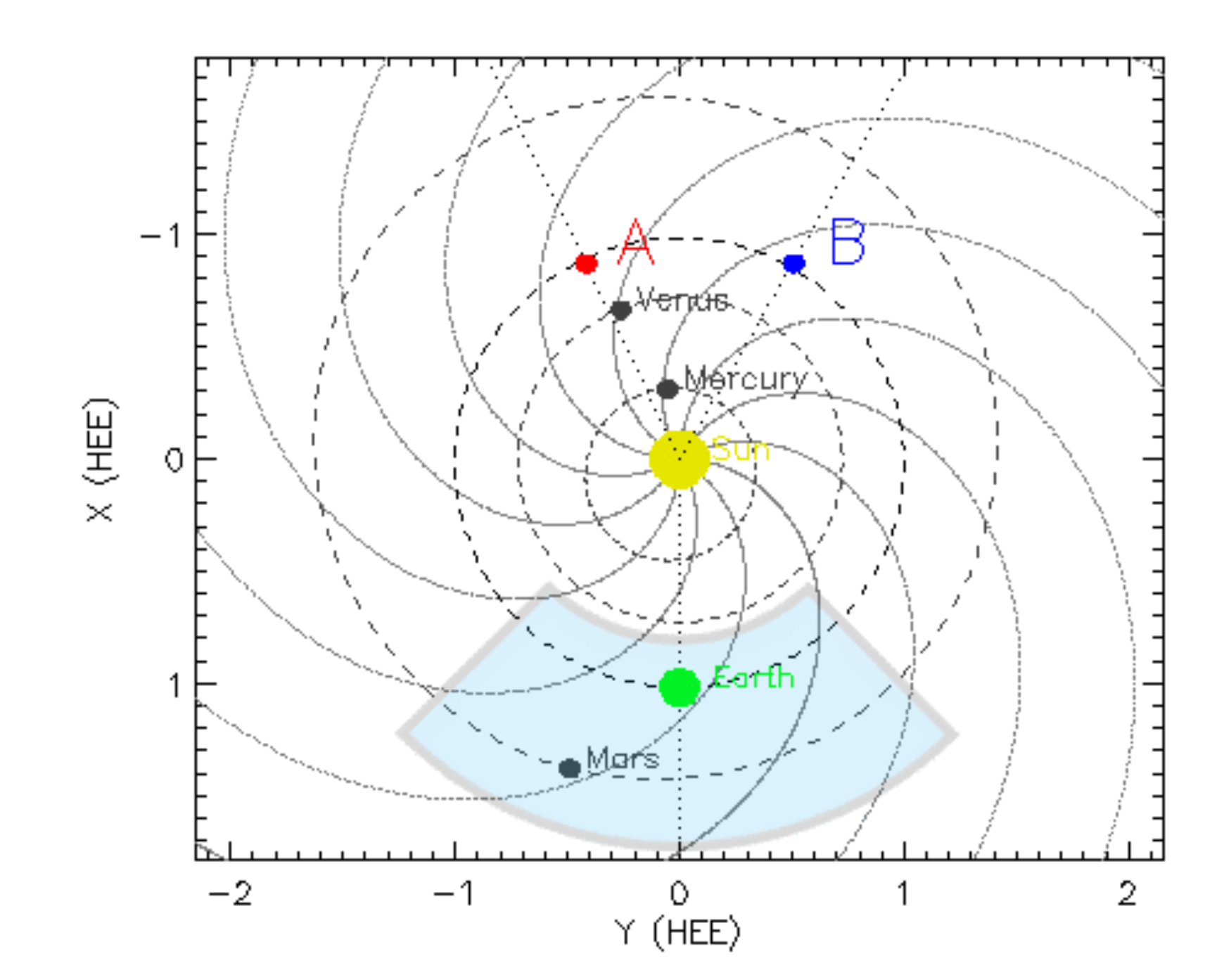}}
\caption{Position of the inner planets and the Solar TErrestrial Relations Observatory \citep[STEREO,][]{Kaiser08} on July 10, 2016. The light blue shaded area marks the angular separation that is referred to as "aligned phase" and spans $\pm 45 \degree$ ahead and behind Earth's heliographic longitude. SIRs and CIRs observed during this period of time are hence called "aligned events". Minimal temporal separation between the measurements of the same structure is given when both planets lie on the same arm of the Parker spiral that is depicted as bent solid gray lines under the approximate SW speed of $400\ \mathrm{km s^{-1}}$. Credits: Stereo Science Center.}
\label{aligned_helio}
\end{figure}

For our study we use in-situ measured plasma and magnetic field data comprising the proton bulk speed, density, and temperature as well as the magnitude of the total $B$. Data used cover 1-min resolution data from the OMNI database \citep{King05} for near-earth in-situ measurements and 45-s resolution data from the MAVEN spacecraft for SW measurements at Mars between November 2014 and November 2018. The MAVEN SW and IMF data in the undisturbed SW have been selected during each orbit taking into account the Martian bow shock structure and the solar wind interactions with Mars \citep{Halekas17}. As the Martian bow shock has significant variability in location, it is possible to get interspersed periods with or without SW coverage. After removing the periods when MAVEN is located within the bow shock, the resolution of MAVEN data is lowered to 270~s on average (ranging from 60--1000~s). We note that for long periods, when MAVEN was not at all in the upstream solar wind, data are fully excluded.  

The catalogs given by \citet{Grandin19} and \citet{Huang19} are used for identification of time ranges covering SIRs and/or CIRs at Earth and Mars, respectively. The time intervals of the events are extracted from these catalogs and applied to the respective spacecraft data to determine the corresponding SW parameter profiles. \cite{Huang19} reports during the period from November 2014 to November 2018 a total number of 126 SIRs that could be clearly identified at Mars. \citet{Grandin19} analyzed from in-situ data at Earth/L1 location a total of 588 SIR/CIR events between January 1995 and December 2017 (101 during November 2014--December 2017). For covering the same time period including the entire data set for Mars, we enlarged the catalog provided by \citet{Grandin19} and we manually identified SIRs/CIRs from December 2017 until November 2018 by inspecting ACE data. In order to restrict the potential SIR/CIR time ranges, we followed a two-sided approach. On the one hand, we related CIRs already identified for that time range at Mars by \citet{Huang19} back to Earth and on the other hand, we manually cross-checked the in-situ data (OMNI) as well as excluded coronal mass ejections identified from the Near-Earth Interplanetary Coronal Mass Ejections list by \citet{CaneRichardson03}.

The back-relating procedure for SIRs/CIRs from Mars to Earth is based on Eq.~\ref{eq1} giving an estimation of the time delay, $\Delta t$, of the occurrence of an HSS between Earth and Mars according to \cite{Opitz2009} with
\begin{equation} \label{eq1}
\Delta t = t_\mathrm{M} - t_\mathrm{E} = \frac{\phi_\mathrm{M} - \phi_\mathrm{E}}{\omega_\mathrm{sun}} + \frac{r_\mathrm{M} - r_\mathrm{E}}{v_\mathrm{SW}}, 
\end{equation}
where $\phi_\mathrm{M} - \phi_\mathrm{E}$ denotes the relative heliographic longitude of the planets, $ \omega_\mathrm{sun} $ the sidereal rotation rate of the Sun, $ r_\mathrm{M} - r_\mathrm{E} $ the relative radial distances of Earth and Mars, and $ v_\mathrm{SW} $ represents the median SW speed for the duration of the HSS at the location of Earth. An HSS is defined as a period over which the SW speed is enhanced as compared to the background SW. In general, the speed profile of an HSS is asymmetric, gradually increasing to its maximum within about one day and then slowly decreasing for several days. In this study, the start of the stream is set to the occurrence of the density peak, the end of the stream is defined when the bulk SW stream drops below $350\ \mathrm{km s^{-1}}$ (approximately corresponding to the minimum speed of an HSS at both planets, cf. Table~\ref{SIRprops}). For streams not reaching this value until the next stream occurs, the minimal value before this next stream is taken as boundary. If, for a given HSS at Mars, there is a corresponding structure detected at Earth (i.e., sharp density peak followed by a rise in and local maximum of the proton bulk speed) lying within the uncertainty boundaries of the expected time delay and not falling into the transit time of a CME, this structure is classified as an SIR at Earth. SIRs are identified as CIRs according to their persistence for at least one solar rotation \citep[see e.g.,][]{Smith76}. In total our data set, which is consistent in time for Earth and Mars, covers 146 SIR/CIR events at Earth and 126 SIR/CIR events at Mars. Note that the lower number of events at Mars might be due to frequent data gaps of the MAVEN SW measurements.

\subsection{Superposed epoch analysis of SIRs} \label{SEAmethod}
For the full data set, we applied a superposed epoch analysis (SEA), which is a statistical means to extract characteristic properties of the individually measured profiles. We follow the original approach given by \citet{Chree13} where the occurrence of a key event embedded in a larger profile is defined as zero epoch time. Here, the zero epoch time is set to be the occurrence of the well-observable and sharp peak\footnote{The density peak in the MAVEN data is found to be more significant compared to the SI peak.} in proton density as it always slightly precedes the SI and drops sharply at the SI \citep[see e.g.,][]{Gosling99,Jian06}. We note that due to MAVEN traversing through the magnetically shielded night side of Mars, we assume for the determination of the zero epoch an uncertainty of about $\pm$2.25 h. Magnitudes derived by SEA in this study present lower boundaries for the actual magnitudes. Individual SW profiles of SIR/CIR events are stacked into an array considering a total time range of 7 d, covering $-$2 d, which is identical to 2 d before the density peak (zero epoch time) and $+$5 d after the occurrence of the density peak. The asymmetry of this interval corresponds to the spatial asymmetry of the combined SIR and HSS structure, being compressed in the rising speed part and rarefacted in the trailing edge. From the resulting array, which shows the most dominant recurring structures from the sample of individual profiles, we derive the median and the $25^\mathrm{th}$ and $75^\mathrm{th}$ percentile (i.e., lower and upper quartile, respectively).

\subsection{Aligned events analysis} \label{AEAmethod}

From a statistical study using data from STEREO and the Solar Dynamics Observatory \citep[SDO][]{Pesnell12}, it is derived that the source regions of SIRs, i.e, CHs, do not show a strong evolution over 2--3 d, especially during low solar activity phases \citep[c.f.][]{Temmer18}. Based on that result we extract a subset of SIR/CIR events for times of small separation angles between Earth and Mars, hence, so-called opposition phases of Mars (i.e., when the Sun and Mars are on the opposite side as observed from Earth). That enables us to study in more detail the radial evolution of SIRs and CIRs from 1 to 1.5~AU with only weak effects of temporal evolution. Figure~\ref{aligned_helio} shows the separation between Earth and Mars for the longitudinal range E45--W45 over which we refer the planets to be ``quasi-aligned,'' corresponding to two time intervals, March until September 2016 as well as April until November 2018. The subset of aligned SIRs/CIRs observed at both planets covers for the first interval 20 events and for the second interval 22 events (see Appendix).

\begin{figure}
\centering
\resizebox{\hsize}{!}{\includegraphics{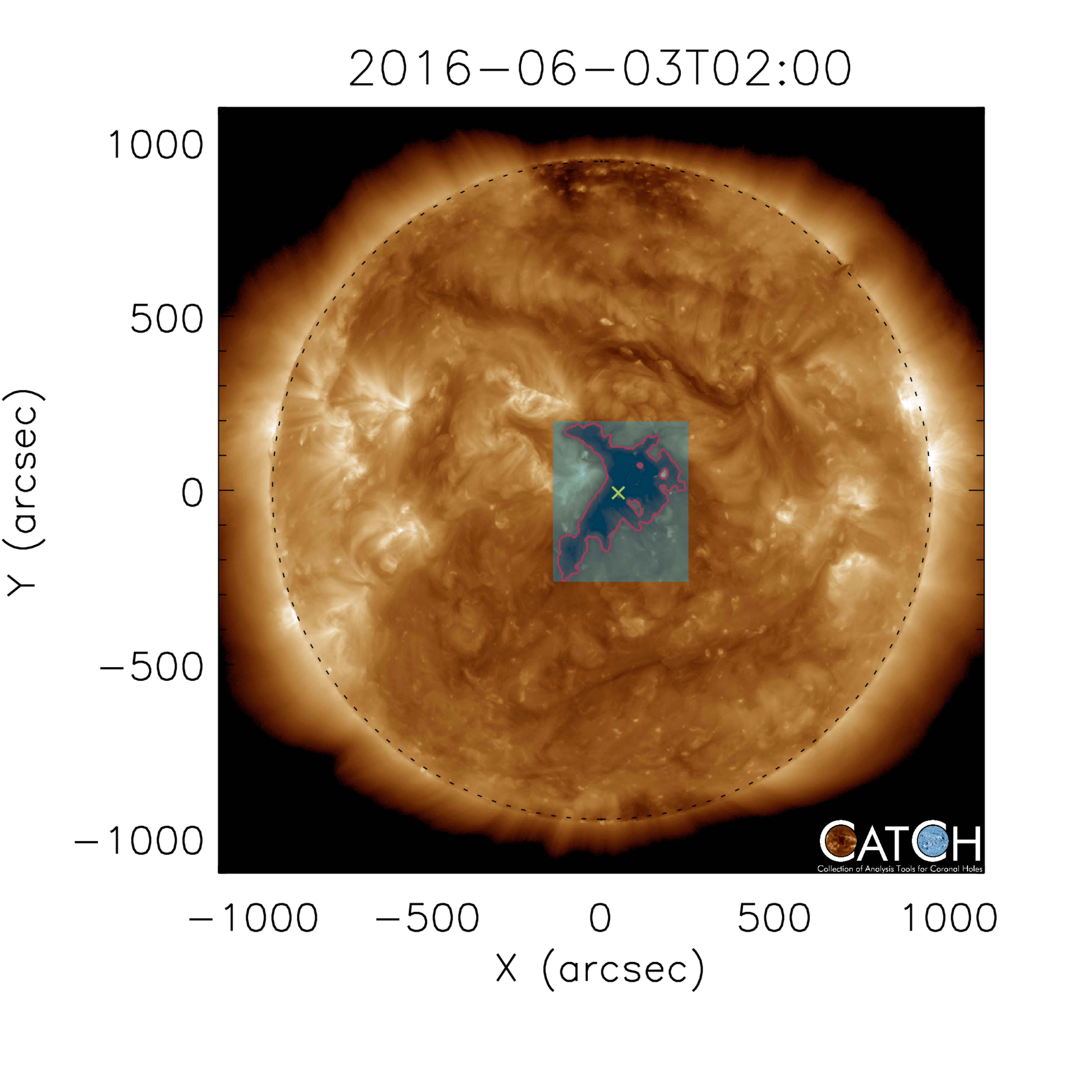}}
\caption{Example of CH extraction using CATCH from June 3, 2016; SDO/AIA 193~\AA. The area of the CH is outlined by a red line marking the boundary obtained by using a threshold of 44\% of the median intensity of the solar disk. The yellow cross denotes the CH center-of-mass, and the horizontal/vertical edges of the light blue shaded box give the maximum longitudinal/latitudinal extent.}
\label{CATCH_expl}
\end{figure}

For a more complete picture we study for the aligned events also the source CH properties. The corresponding CHs \citep[identified via the central meridional passage 2--3 d before the peak in density, see e.g.,][]{Vrsnak07} are extracted using SDO/EUV data and the Collection of Analysis Tools for Coronal Holes \citep[CATCH; see][]{Heinemann19}. CATCH is a supervised intensity-threshold extraction method that is modulated by the intensity gradient at the CH boundary, which also covers uncertainty estimations of area and boundaries. From that we derive for each SIR the associated CH area, its latitudinal and longitudinal extent, as well as the geometrical center-of-mass (CoM) in latitude and longitude. An example of a CH and its parameters extracted with CATCH is given in Figure~\ref{CATCH_expl}.  

For deriving the correlations between SW and CH parameters we use the bootstrap method due to the low number of data points (in total 42). We apply 1,000 repetitions from which 1,000 statistical samples are created. From that we calculate the Spearman mean correlation coefficient at an 80\% confidence level \citep{efron1979_bootstrap,efron93_bootstrap}.

\section{Results} \label{results}
\subsection{Superposed epoch analysis} \label{SEAresults}

\begin{figure*}
\begin{subfigure}[b]{.5\textwidth}
\includegraphics[width=0.9\textwidth]{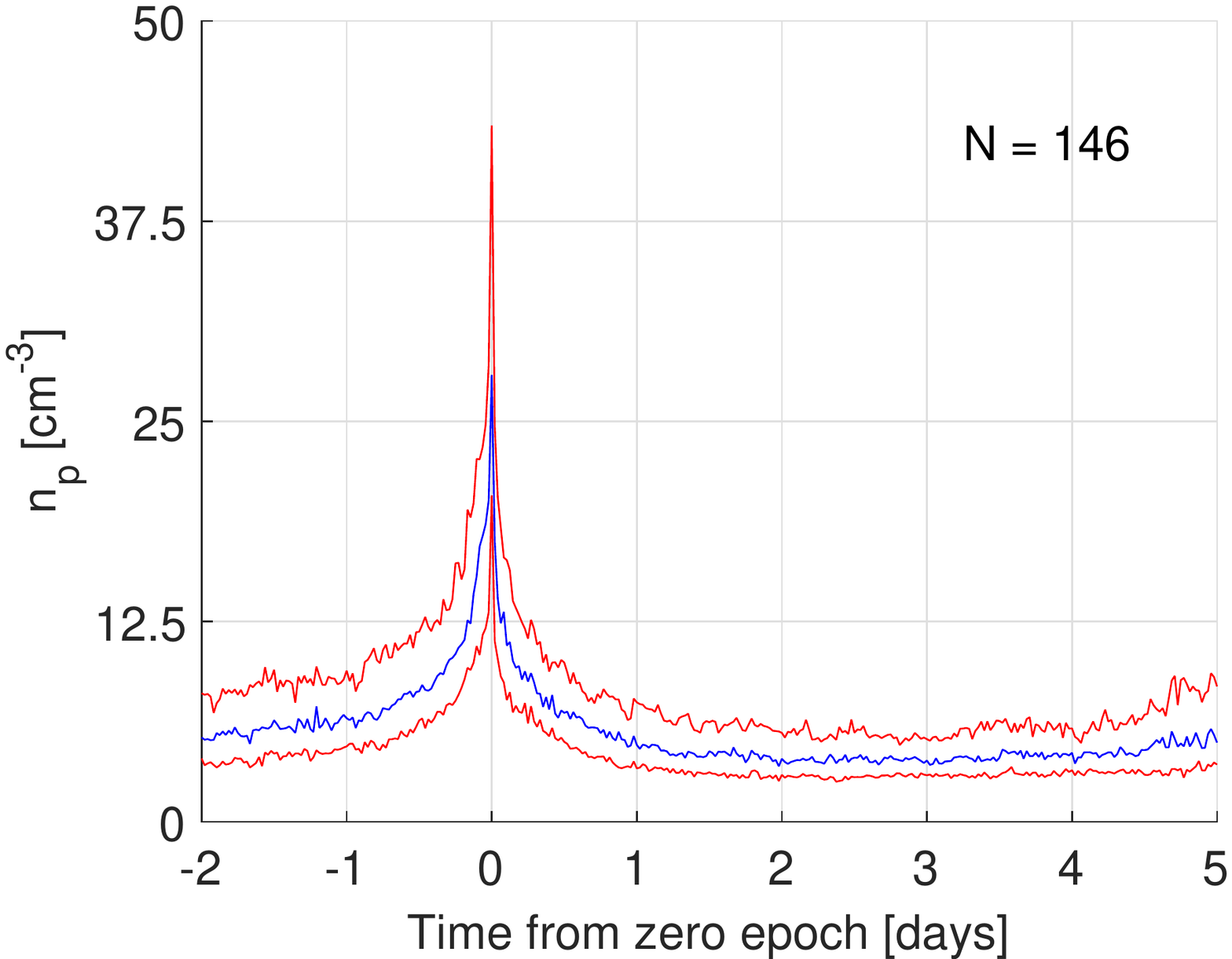}
\end{subfigure}%
\begin{subfigure}[b]{.5\textwidth}
\includegraphics[width=0.9\textwidth]{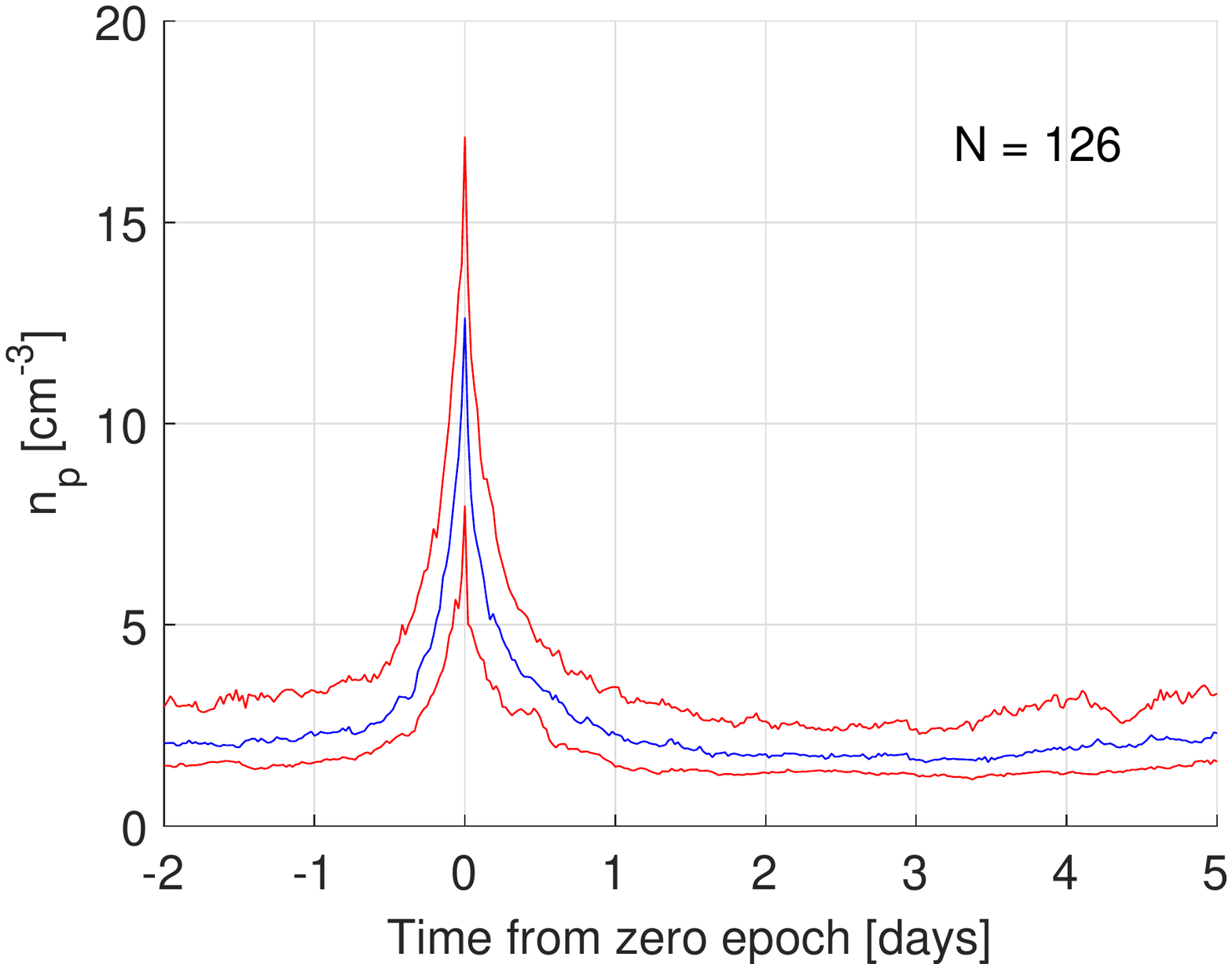}
\end{subfigure}
\caption{Superposed epoch analysis for solar wind density at Earth (left panel) and Mars (right panel). The maximum density of every event is shifted to 0 on the x-axis. The blue curve represents the median, the upper and lower red curves represent the upper and lower quartile, respectively. Mind the different scaling of the y-axes of the two plots.}
\label{nmed}
\end{figure*}

\begin{figure*}
\begin{subfigure}[b]{.5\textwidth}
\includegraphics[width=0.9\textwidth]{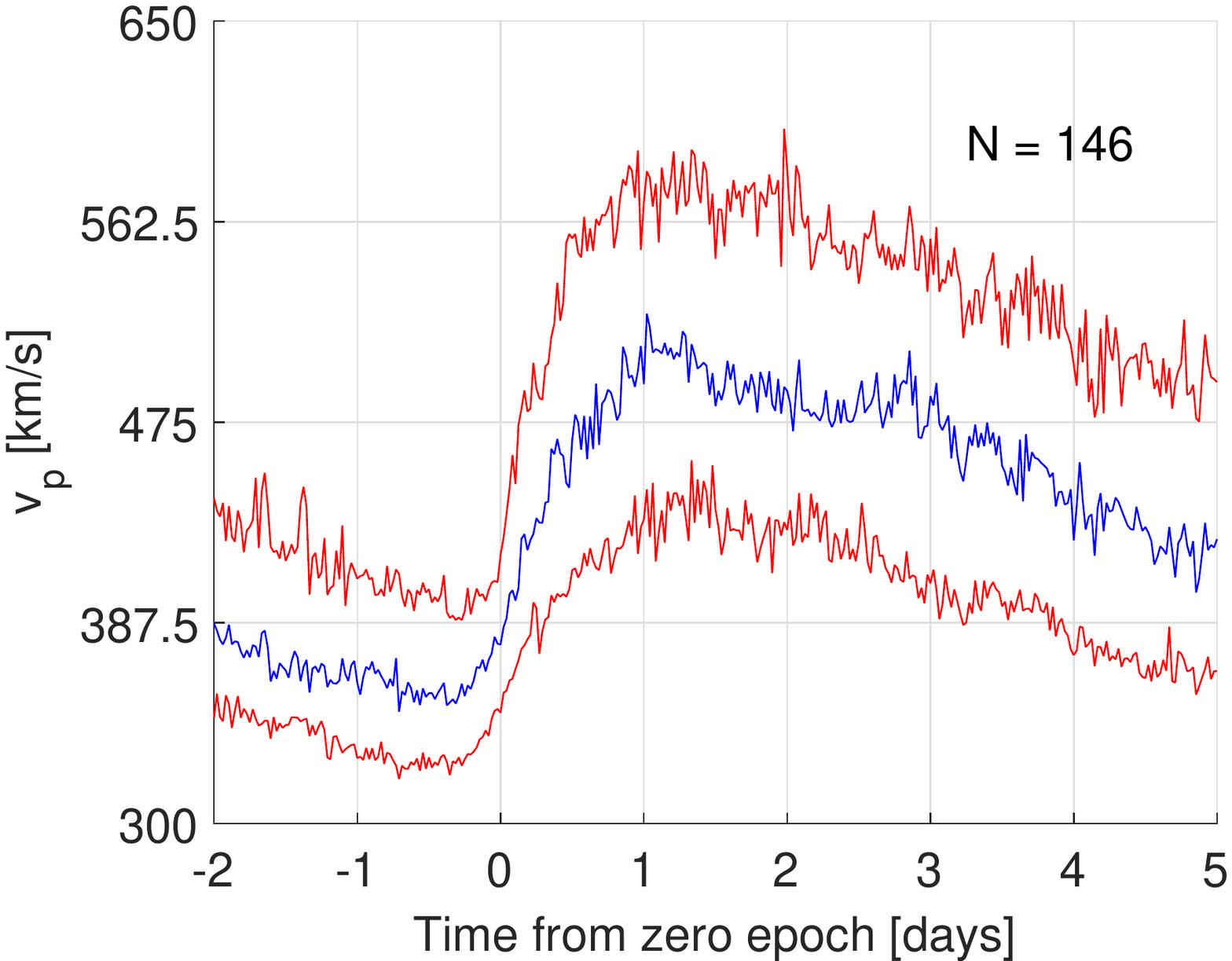}
\end{subfigure}%
\begin{subfigure}[b]{.5\textwidth}
\includegraphics[width=0.9\textwidth]{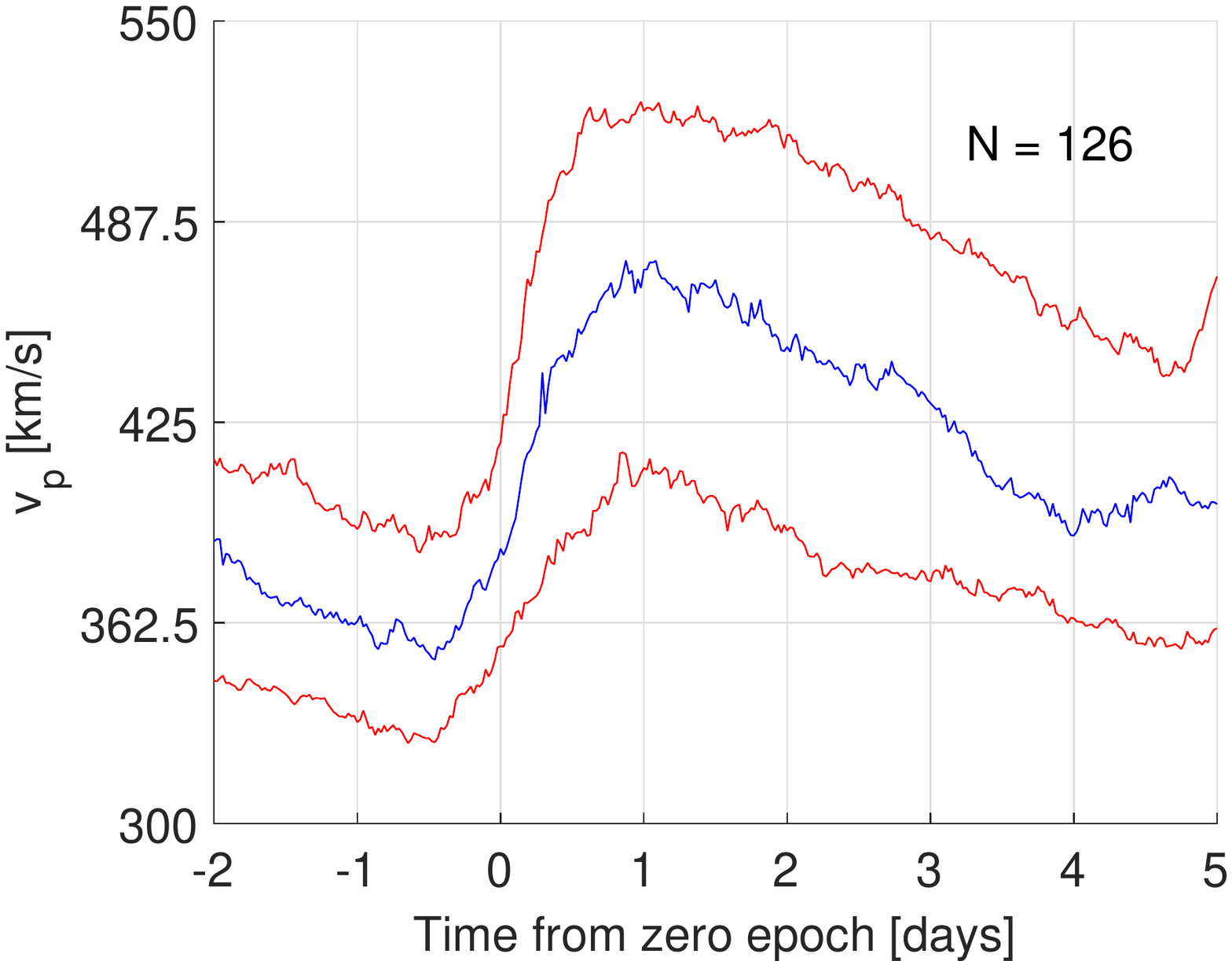}
\end{subfigure}
\caption{Same as Fig. \ref{nmed} for the bulk solar wind speed.}
\label{vmed}
\end{figure*}

\begin{figure*}
\centering
\begin{subfigure}[b]{.5\textwidth}
\includegraphics[width=0.9\textwidth]{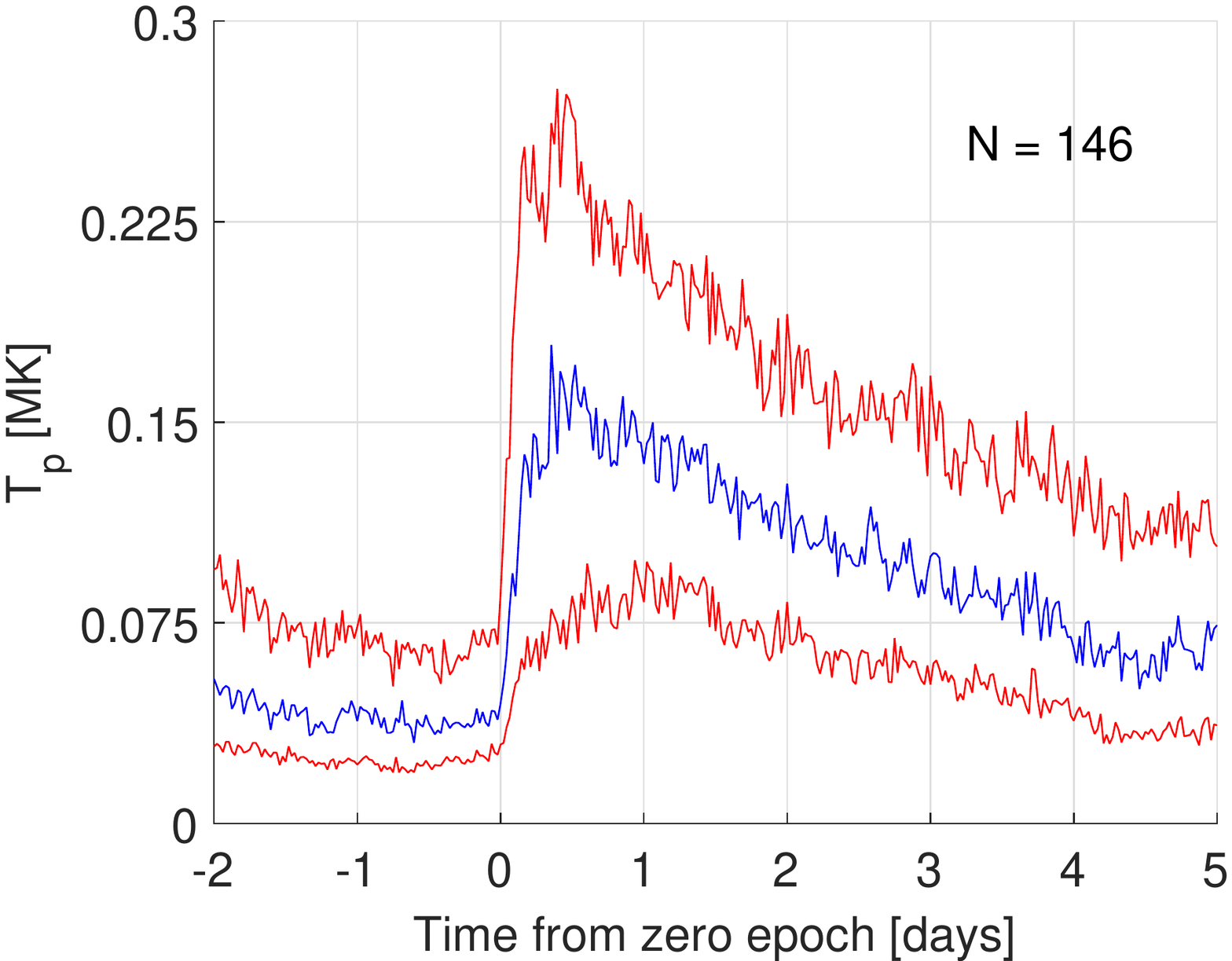}
\end{subfigure}%
\begin{subfigure}[b]{.5\textwidth}
\includegraphics[width=0.9\textwidth]{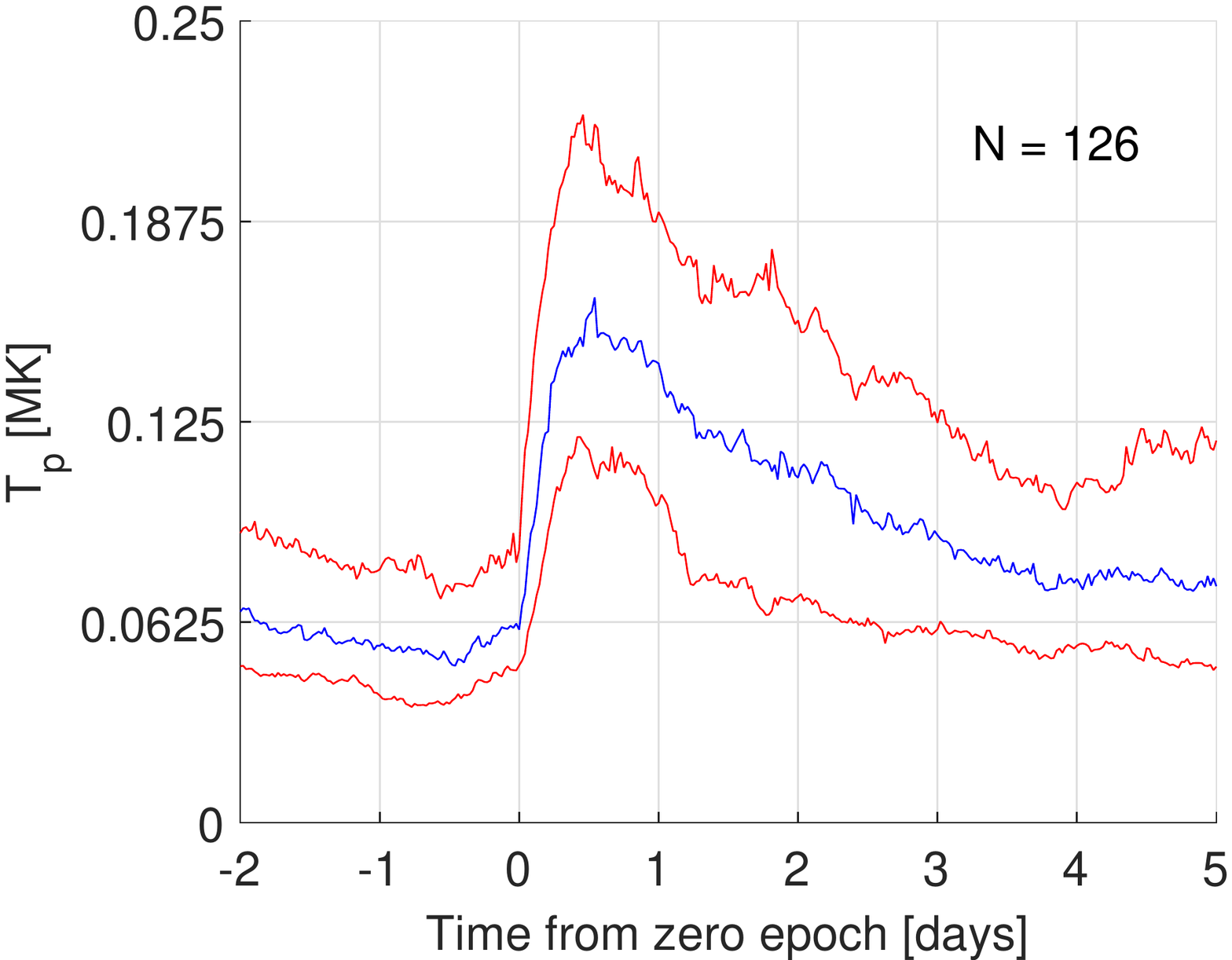}
\end{subfigure}
\caption{Same as Fig. \ref{nmed} for the proton temperature.}
\label{tmed}
\end{figure*}

\begin{figure*}
\begin{subfigure}[b]{.5\textwidth}
\includegraphics[width=0.9\textwidth]{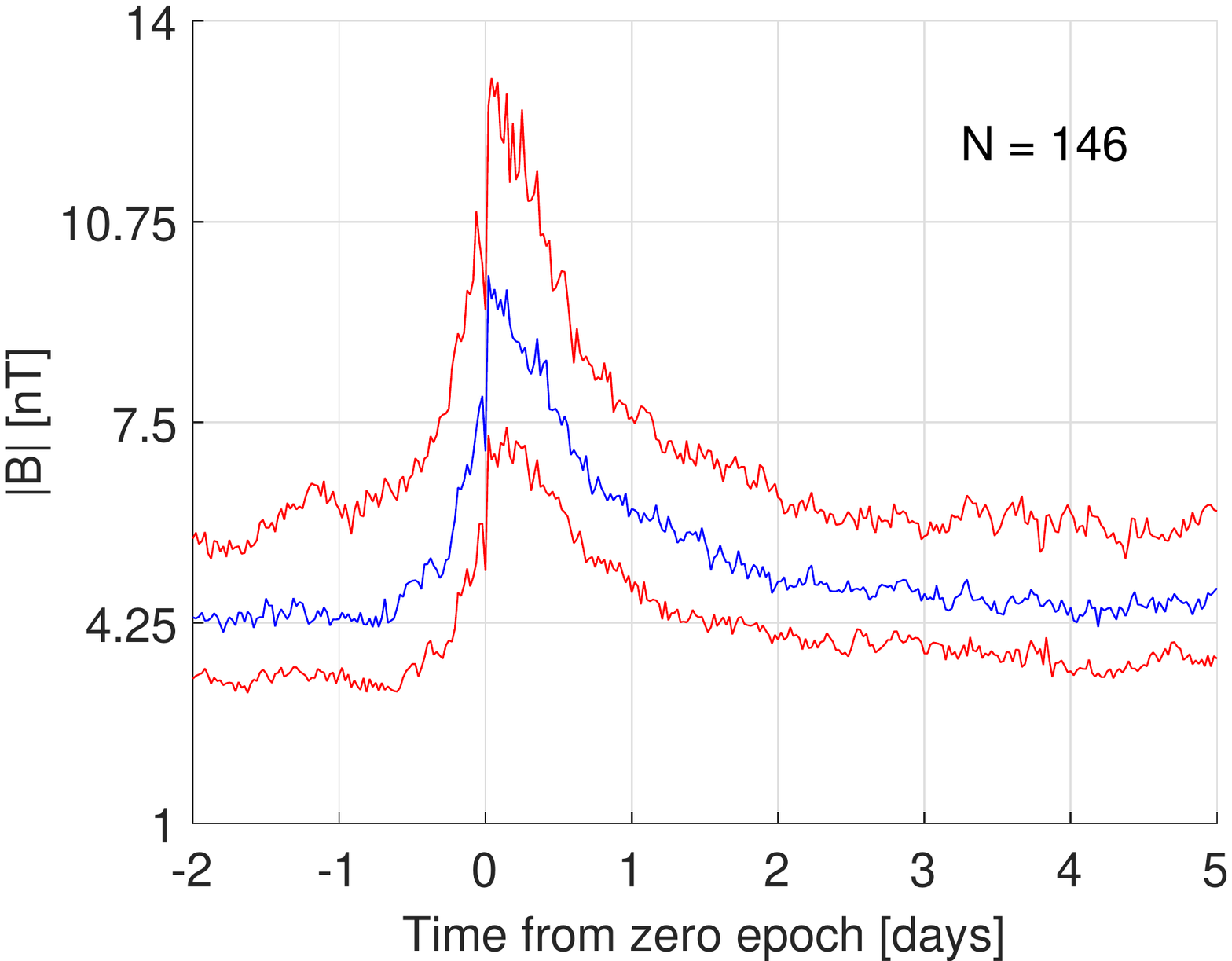}
\end{subfigure}%
\begin{subfigure}[b]{.5\textwidth}
\includegraphics[width=0.9\textwidth]{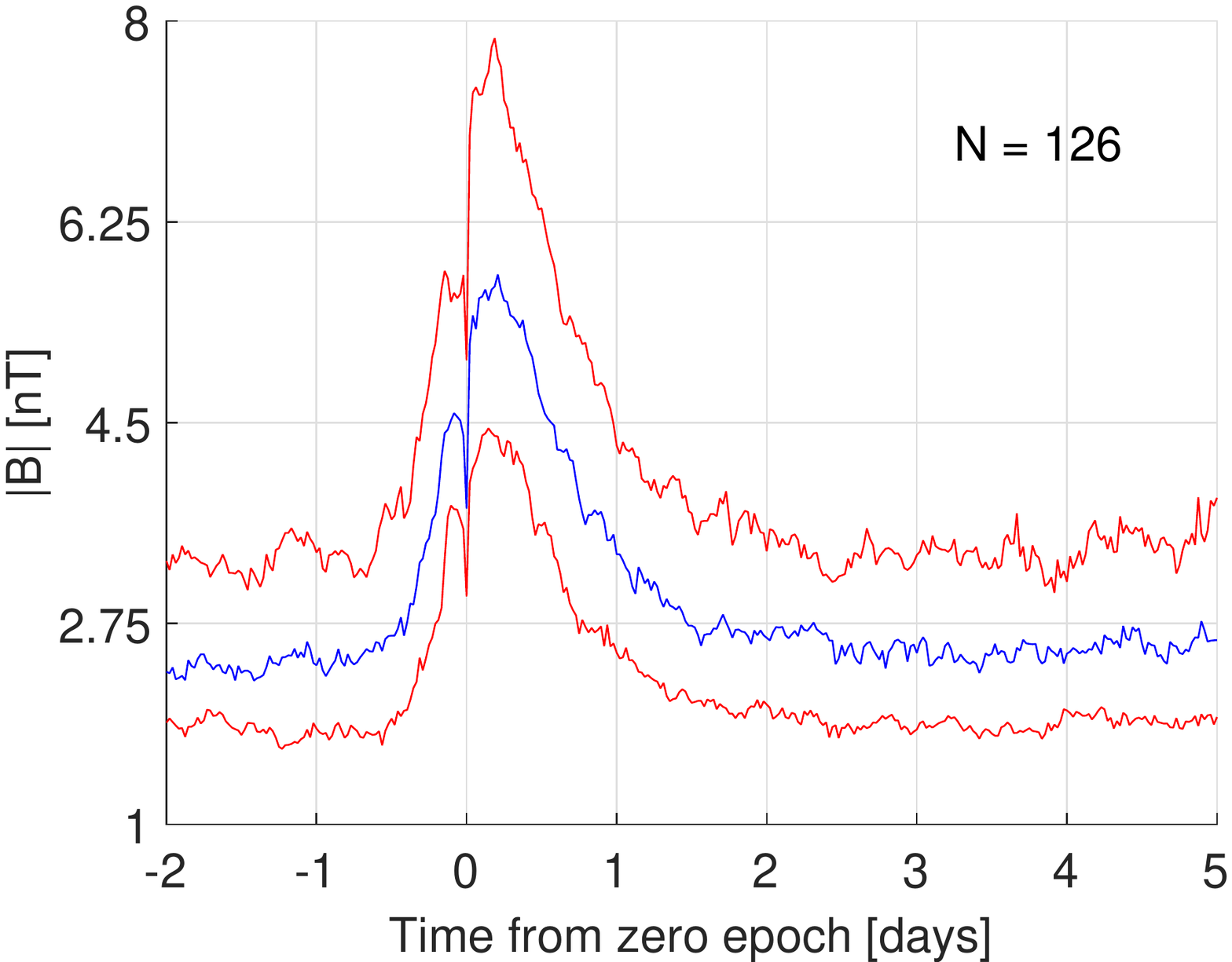}
\end{subfigure}
\caption{Same as Fig. \ref{nmed} for the interplanetary magnetic field magnitude.}
\label{Babsmed}
\end{figure*}

\begin{figure*}
\begin{subfigure}[b]{.5\textwidth}
\includegraphics[width=0.9\textwidth]{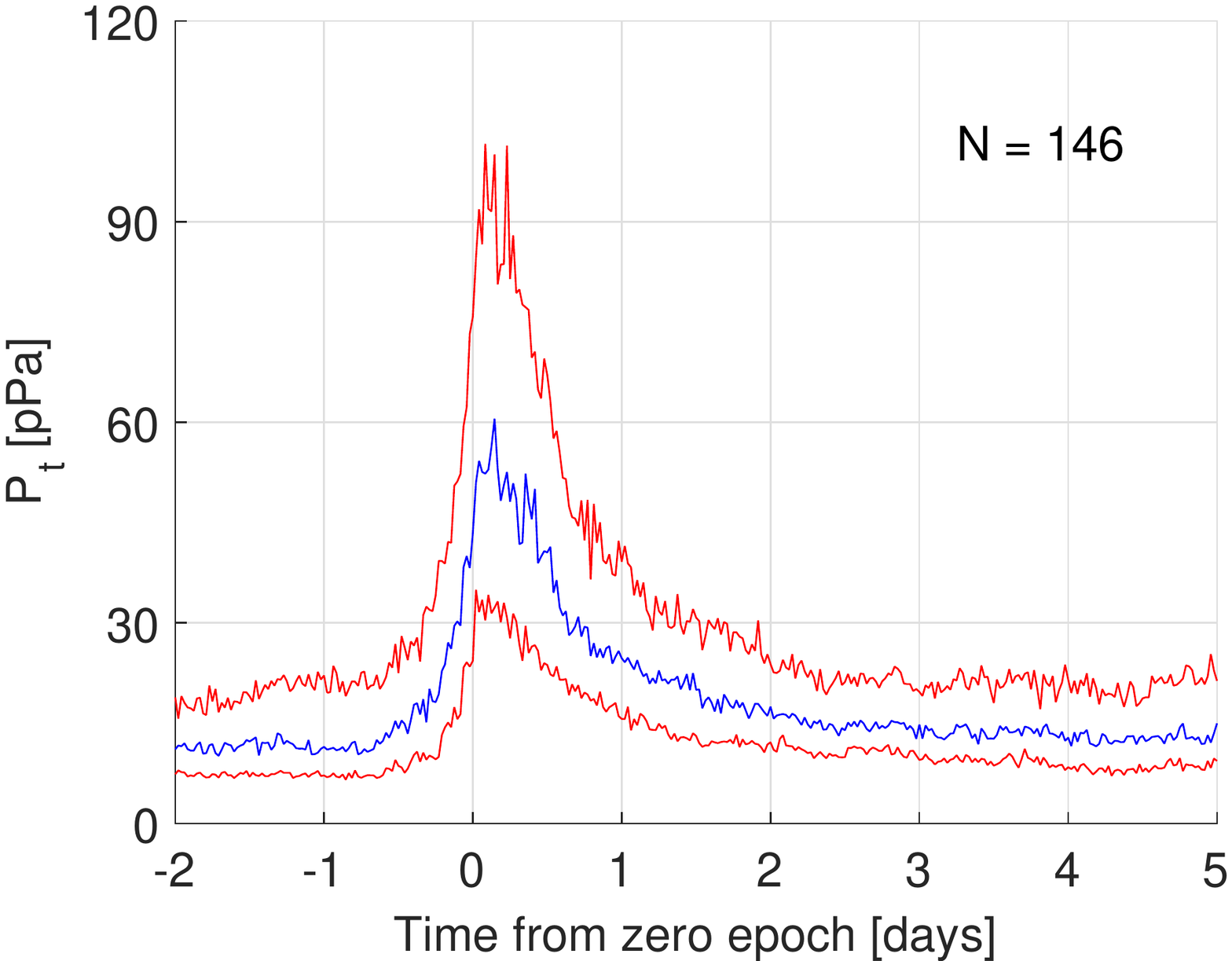}
\end{subfigure}%
\begin{subfigure}[b]{.5\textwidth}
\includegraphics[width=0.9\textwidth]{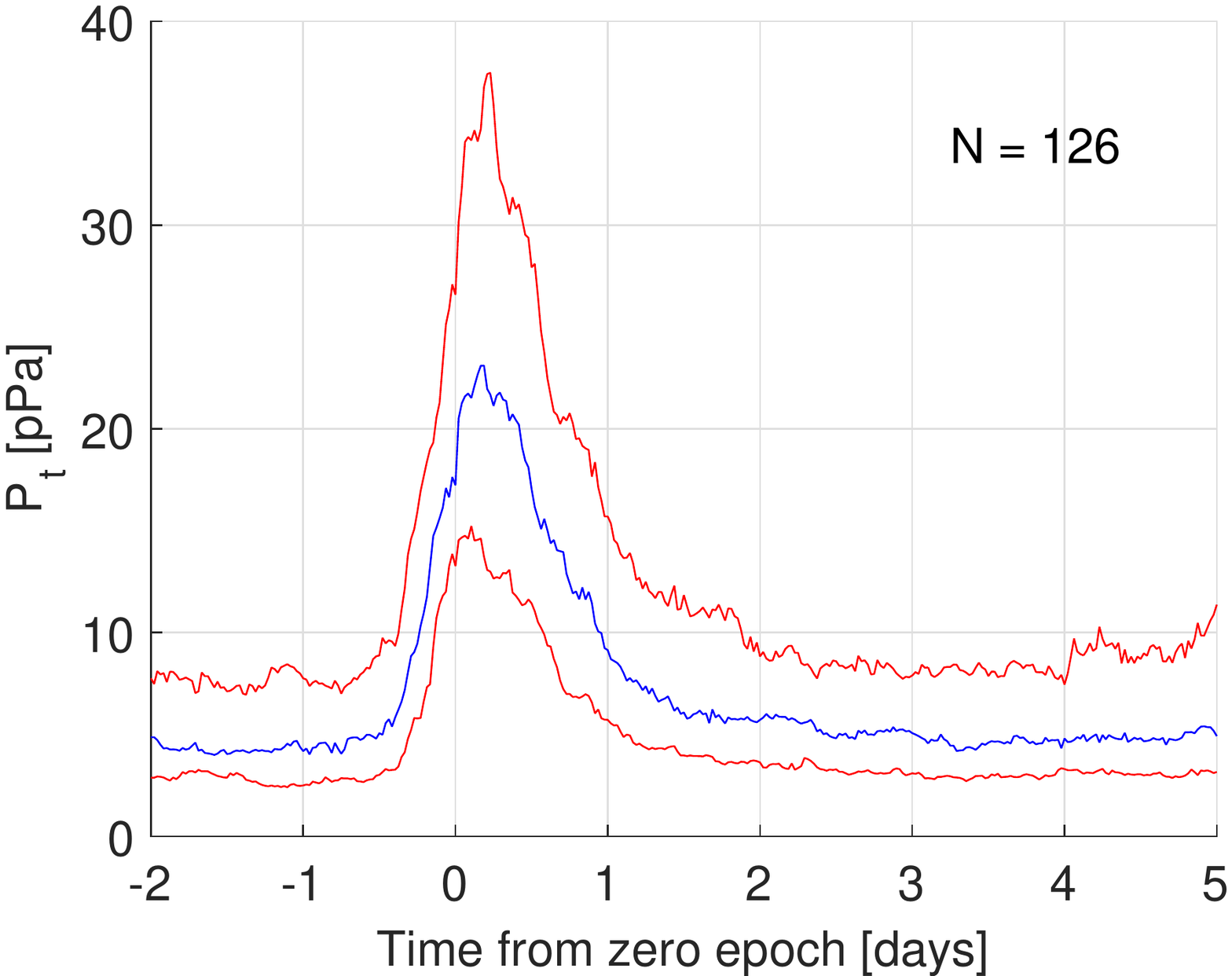}
\end{subfigure}
\caption{Same as Fig. \ref{nmed} for the total perpendicular pressure.}
\label{Ptmed}
\end{figure*}

The SEA was performed for the full sample of events showing SW parameters of SIRs/CIRs separately measured at Mars and Earth including bulk speed, proton density and temperature as well as the components of the magnetic field and the total perpendicular pressure. The resulting profiles with the median as well as the upper and lower quartile are given in Figs. \ref{nmed} to \ref{Ptmed}.

In relation to the zero epoch time, for the number density we derive a gradual increase from $-$1 d that strongly steepens at about $-$0.3 d until reaching the density peak at zero epoch time (cf.\,Fig. \ref{nmed}). For Earth, the median of the peak number density is $28.0\ \mathrm{cm^{-3}}$, while for Mars it is $12.5\ \mathrm{cm^{-3}}$. After the density peak a strong decrease in density is observed that lasts at both heliospheric locations about 24 h but is slightly steeper at Earth than at Mars.  

For the bulk speed we derive for both Earth and Mars a characteristic drop in the profile at about $-$0.5 d, followed by a strong increase of the HSS slightly before the zero epoch time (cf.\,Fig. \ref{vmed}). This ``turning point'' \citep[cf.][]{richter86} is located for Earth slightly closer to the zero epoch compared for Mars distance ($-$0.38 d vs. $-$0.46 d). The maximum of the median bulk speed for Earth is $522\ \mathrm{km s^{-1}}$ reached at $+$1.02 d after the median density peak, and for Mars we derive a maximum value of $475\ \mathrm{km s^{-1}}$ which is reached at $+$0.88 d from zero epoch. 

The median plasma temperature increases strongly after zero epoch time, having values that remain relatively low prior to the occurrence of the median density peak (cf.\,Fig. \ref{tmed}). At the median density peak, the onset of a rise that is much steeper than that of the median bulk speed, is clearly visible, leading to a maximum of $1.8\times 10^{5}\ \mathrm{K}$  at $+$0.35 d for Earth and to $1.6 \times 10^{5}\ \mathrm{K}$ at $+$0.54 d for Mars distance. Pre-HSS levels are not reached for neither of the two planets after 5 d.

The magnitude of the IMF also shows notable properties, as can be seen in Fig. \ref{Babsmed}. The profiles at Earth and Mars start to rise at $-$1 d of zero epoch time, peaking shortly after the maximum median density. Almost simultaneously at the occurrence of the median density peak, a shock-like drop in magnetic field strength can be clearly observed, for Mars even more pronounced than for Earth. This drop is followed by a steep, nearly shock-like increase until the peak values are reached, $9.9\ \mathrm{nT}$ at Earth and $5.8\ \mathrm{nT}$ at Mars. The median IMF strength peaks shortly after the median density peak, at $+$0.02 d for Earth and $+$0.21 d for Mars. Pre-HSS levels are reached after about $+$3 d in the case of Mars, and about $+$4 d in the case of Earth.

The median total perpendicular pressure is given in Fig. \ref{Ptmed}. This profile combines the bulk density, magnetic field magnitude and temperature \cite[$P_\mathrm{t} = \frac{B^{2}}{2\mu_{0}} +  \Sigma(nkT_\mathrm{perp})$; cf.][]{Jian06}. Accordingly, a rise can be seen from $-$1 d onward, being steepest from the occurrence of the median density peak to the maximum of the profile. The profile for Earth-bound data peaks at $+$0.15 d epoch time with $59.9\ \mathrm{pPa}$. For Mars, the peak value of the median of $P_\mathrm{t}$ is reached at $+$0.17 d epoch time with $23.1\ \mathrm{pPa}$.

Table \ref{SIRprops} summarizes for the plasma and magnetic field properties of the entire sample of SIR/CIR events from the SEA study the resulting minimum and maximum values, as well as the time difference in the peak value relative to zero epoch. 

\begin{table}
\caption{Median plasma and magnetic field properties of SIRs at Earth and Mars together with the time difference (in days) of the maximum values with respect to zero epoch, which is where the peak in density occurs.}
\label{SIRprops}
\centering
\begin{tabular}{c c c c c}
\hline \hline
Parameters & 1 AU & 1.5 AU & $\Delta$t$n_{E}$ & $\Delta$t$n_{M}$ \\
\hline
$v_\mathrm{max}\ [\frac{\mathrm{km}}{\mathrm{s}}]$ & 627 & 505 & 1.02 & 0.88 \\
$v_\mathrm{min}\ [\frac{\mathrm{km}}{\mathrm{s}}]$ & 337 & 356 & $-$0.38 & $-$0.46\\
$n_\mathrm{max}\ [\mathrm{cm^{-3}}]$ & 28.0 & 12.5 & 0. & 0.\\
$B_\mathrm{max}\ [\mathrm{nT}]$ & 14.5 & 7.5 & 0.02 & 0.21\\
$T_\mathrm{p,max}\ [\mathrm{MK}]$ & 0.50 & 0.23 & 0.35 & 0.54\\
$P_\mathrm{t,max}\ [\mathrm{pPa}]$ & 116.0 & 32.2 & 0.15 & 0.17 \\
\hline
\end{tabular}
\end{table}

\subsection{Aligned events analysis} \label{AEAresults}

In the Appendix, Tables \ref{AEA_table2016} and \ref{AEA_table2018} list start and end times, together with the properties of all aligned SIR/CIR events detected between March and September 2016 (20) as well as April and November 2018 (22). The occurrence of each SIR is numbered in chronological order and, if recurrent, also the corresponding CIR is numbered (the CIR keeps the number over its lifetime, facilitating allocation in the table). In summary, the list covers 42 SIR events out of which 8 are recurrent over one to five solar rotations. In addition to the plasma and magnetic field peak values, as derived for each SIR, the information of an associated shock type (fast forward or reverse) as well as the separation angles between Earth and Mars are given.

\begin{figure*}
\centering
\includegraphics[width = 0.8\textwidth]{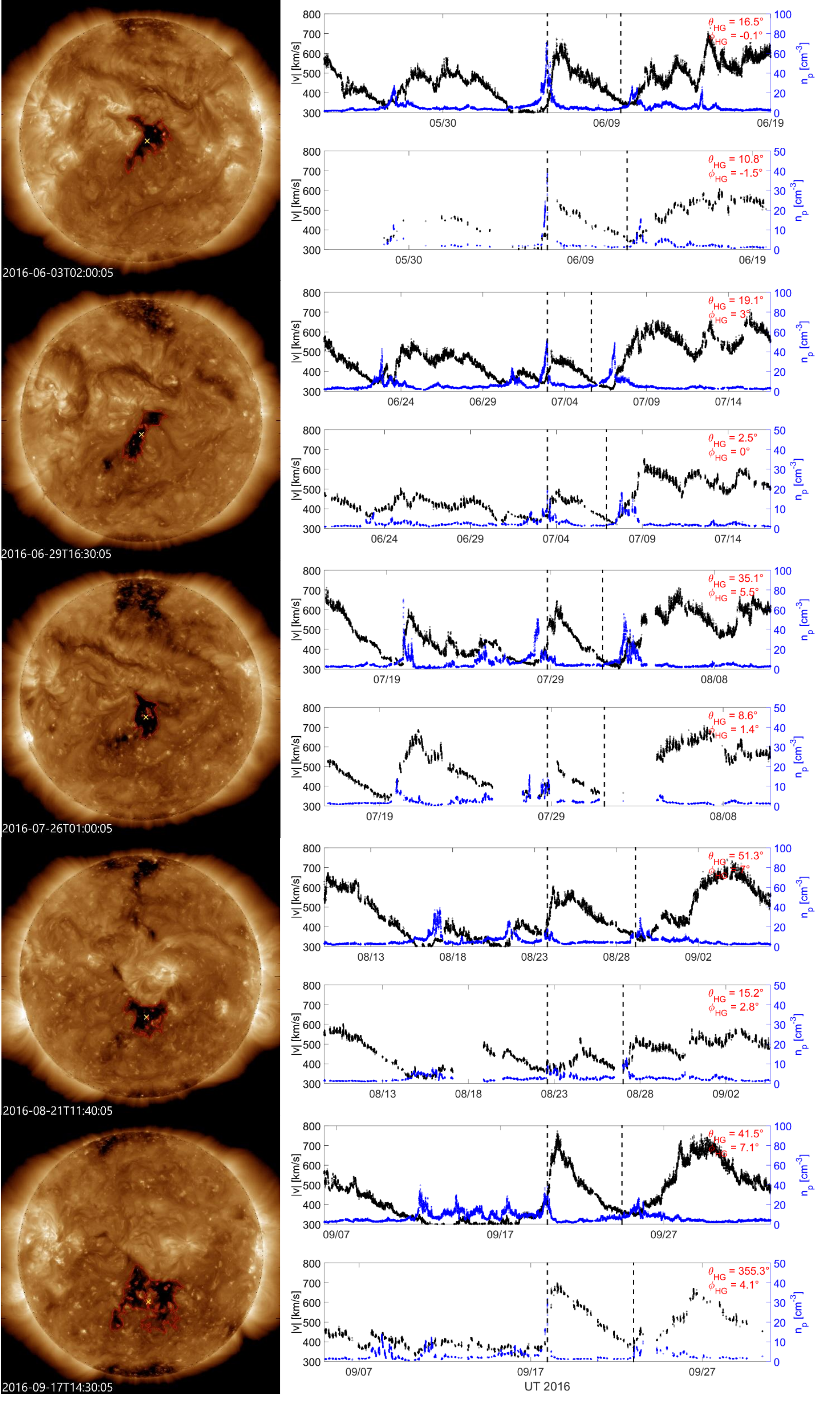}
\caption{SDO/AIA 193 \AA\ images and in-situ SW plasma measurements for CIR no.\,2 recurring five times. \textit{Left:} EUV images showing CHs extracted via CATCH and identified as the origin of HSS measured in-situ. \textit{Right:} In-situ SW bulk speed (red) and density (blue) measured by ACE/WIND (panels in an odd sequence) and MAVEN (panels in an even sequence). The density peak and the end time of the event are marked by black dashed vertical lines. In the top right corner of the profiles the heliographic longitude, $\Theta_{\mathrm{HG}}$, and the heliographic latitude $\phi_{\mathrm{HG}}$, of the planet is given.}
\label{CIR2}
\end{figure*}

Figure~\ref{CIR2} shows an example for the evolution of a CIR over its lifetime covering five solar rotations (given in Table~\ref{AEA_table2016} as SIR no.\,5, 10, 14, 17, 20 and CIR no.\,2). The left panels give the EUV SDO/AIA image data with the corresponding CH extracted using CATCH. The right panels give the in-situ measured solar wind plasma profiles for the proton bulk speed and proton density at Earth (top) and Mars (bottom). The black dashed vertical lines in each profile mark the density peak and the end time of the event, from which we derive the duration of the stream. For deriving the relation of the CH position on the Sun and the stream properties observed at the planet, we note in each profile the heliographic longitude, $\Theta_{\mathrm{HG}}$, and the heliographic latitude $\phi_{\mathrm{HG}}$ of the planet. From this the co-latitude is obtained. Corresponding Figures for CIRs no.\,1 and 3--8 are presented in the Appendix.

\begin{figure*}
\begin{subfigure}[b]{.5\textwidth}
\includegraphics[width=1.\textwidth]{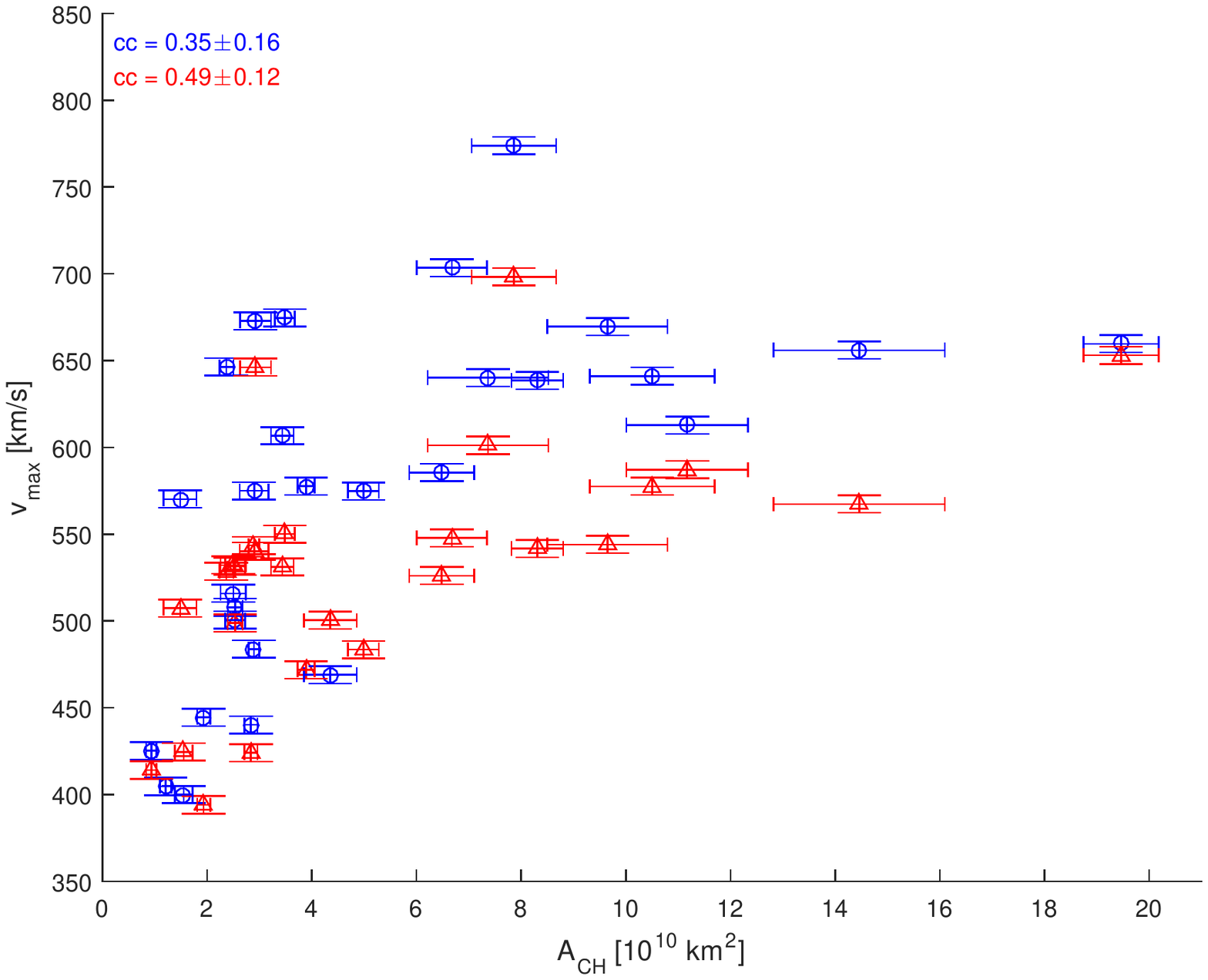}
\end{subfigure}
\begin{subfigure}[b]{.5\textwidth}
\includegraphics[width=1.\textwidth]{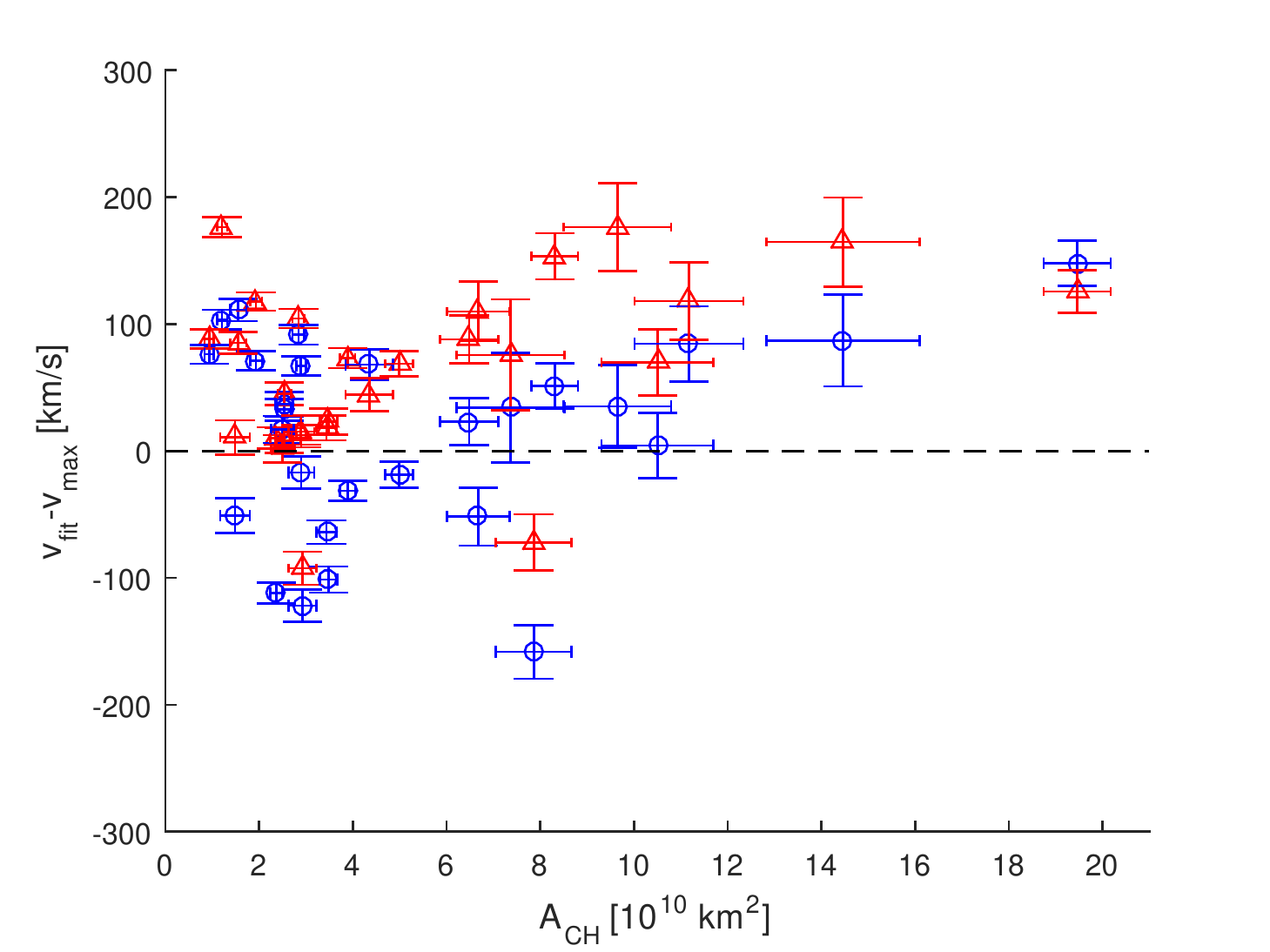}
\end{subfigure}
\caption{{Left:} Maximum in-situ measured SW speed versus CH area. {Right:} Absolute difference between the fitted value and the measurements, with a black dashed line indicating zero difference. The blue circles denote data points related to Earth, the red triangles denote data points related to Mars. The respective colors are also used for the correlation coefficient values as given in the legend. The error of the CH area scales with the uncertainty of extraction, for instance, CHs that lie close to the solar limb have a larger error. The error of the measured maximum bulk velocity was set to $\pm5\ \mathrm{kms{-1}}$.}
\label{cc_area_v_fit}
\end{figure*}

Most similar profiles of SW speed and density measured at Earth and Mars are obtained in the frames 2 and 3 in Fig.~\ref{CIR2} (CH at June 29, 2016 and July 26, 2016). For those times the difference in heliographic longitude is $\sim15-25 \degree$, meaning that the two planets lie on the same arm of the Parker spiral (assuming a solar wind speed of 400~km/s) and the temporal evolution of the stream is therefore minimal. While the density between Earth and Mars drops by a factor of $\sim2$, the speed decreases only slightly over that distance. The difference in heliographic latitude ($\sim3-4 \degree$) seems to have little influence on that results. However, the second to last CH configuration observed on August 21, 2018 produced quite different streams at both heliospheric distances. The bulk speed measured by MAVEN shows a slow increase, peaking only $\sim$2.5 d after the density peak. In this case, the difference in heliographic latitude is maximal for the observed CIR, reaching 4.2\degree. Possibly, the planets are located on different sides of the heliospheric current sheet (HCS) at this time, due to both relatively large latitudinal and longitudinal separation. As the CH area and location on the Sun changes with each rotation, also the speed and density profile varies, showing in general the trend of increasing CH area and increased SW speed as clearly shown in the left panel of Fig.~\ref{cc_area_v_fit}.

\begin{figure*}
\centering
\includegraphics[width=0.7\textwidth]{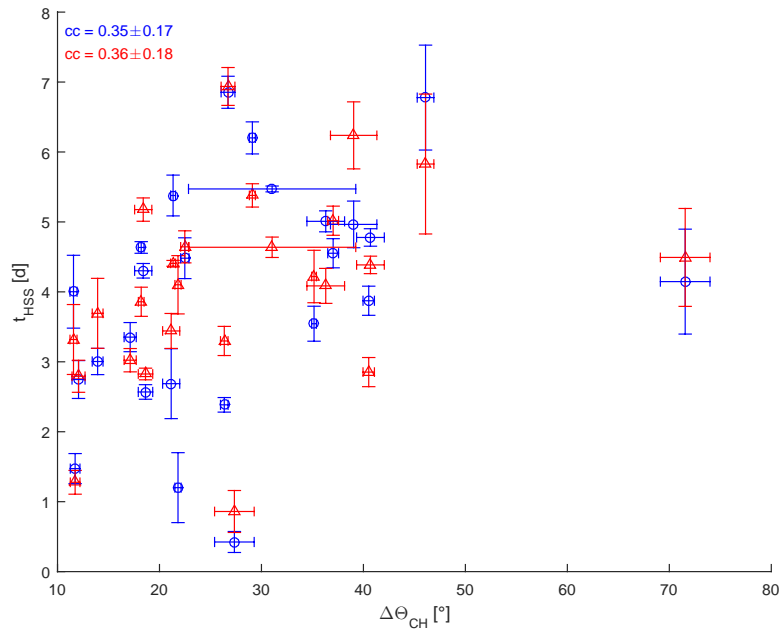}
\caption{SW HSS duration versus CH longitudinal extent. Same color scheme is used as for Figure~\ref{cc_area_v_fit}. The error of stream duration depends on the uncertainty of defining the stream end, for example, if the lower boundary of $350~\mathrm{km s^{-1}}$ is not reached before the occurrence of another SIR/CIR. The error of CH longitudinal extent again grows the closer it is located at the solar limb.}
\label{cc_width_SD}
\end{figure*}

\begin{figure*}
\begin{subfigure}[b]{.5\textwidth}
\includegraphics[width=1.\textwidth]{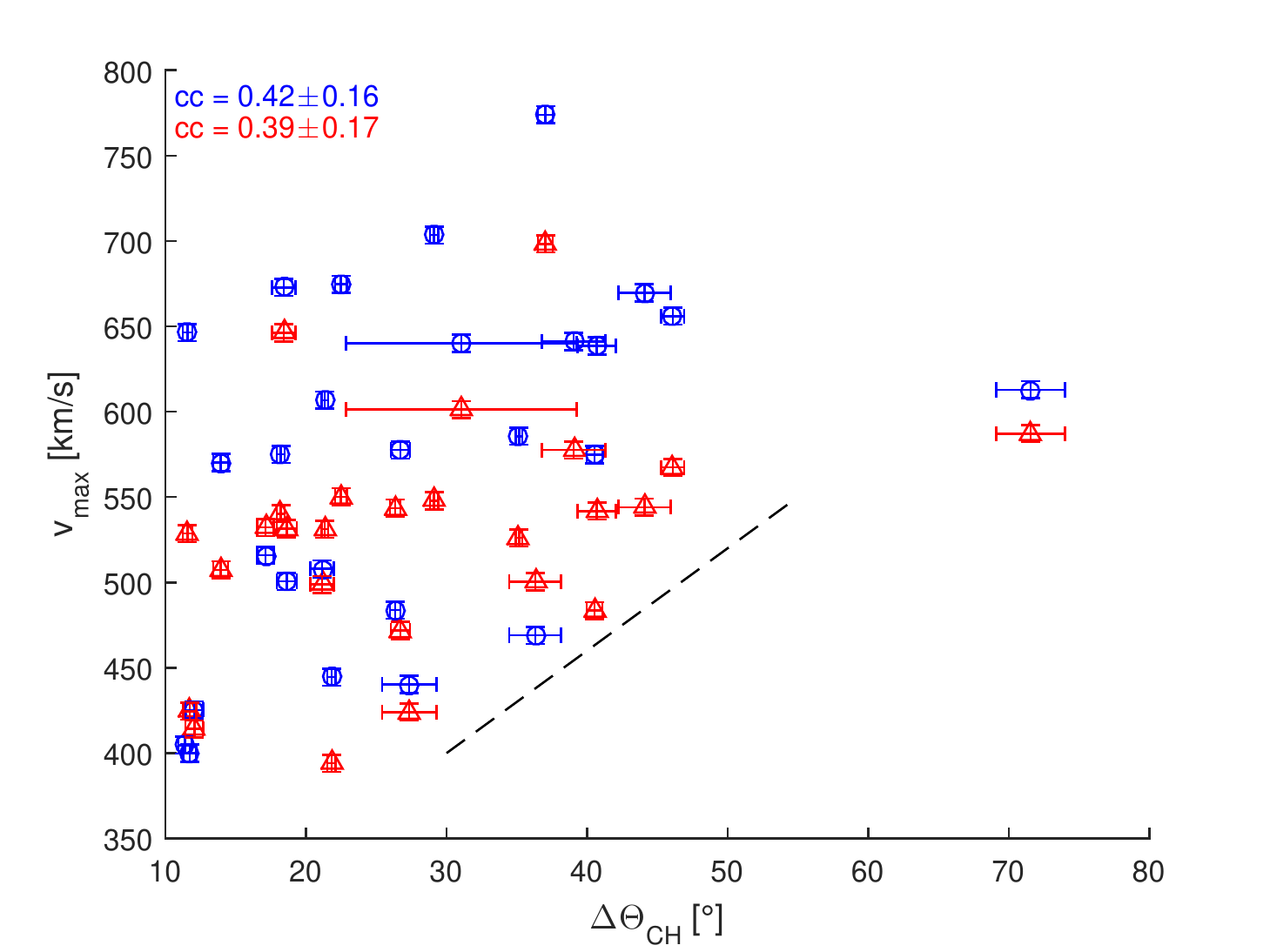}
\end{subfigure}
\begin{subfigure}[b]{.5\textwidth}
\includegraphics[width=1.\textwidth]{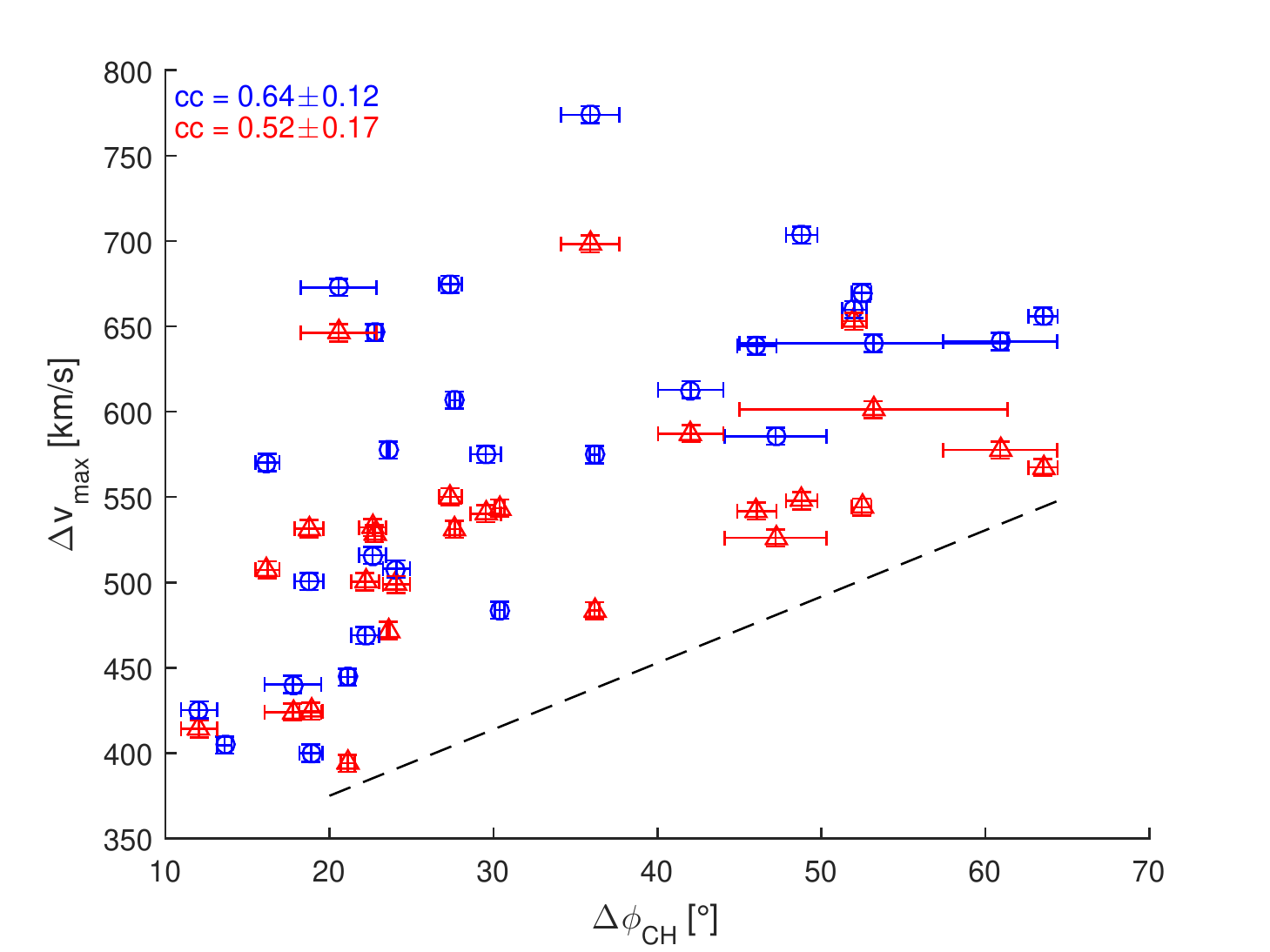}
\end{subfigure}
\caption{{Left:} SW maximum speed versus CH longitudinal extent. {Right:} SW maximum speed versus CH latitudinal extent. Same color scheme is used as for Figure~\ref{cc_area_v_fit}. Again, greater error applies to CHs lying close to the limb or polar regions.}
\label{cc_CH_speed}
\end{figure*}

For deriving more details on the characteristics of SIRs/CIRs at Earth and Mars in relation to their solar sources, we performed correlation analyzes from which the results are shown in Figs. \ref{cc_area_v_fit} to \ref{cc_CH_speed}. We first investigate the well-known empirical relation between CH area and maximum SW speed of a SIR/CIR at 1~AU \citep{nolte76}. In more recent studies, \citet{Hofmeister18} found for a dataset covering CHs over 2010--2017 a linear dependence of the maximum SW speed on the co-latitude of the CHs and the CH area valid for 1~AU distance. The relation is given by Eq.~\ref{vfit},

\begin{equation} \label{vfit}
v_{\mathrm{fit}}(\mathrm{km s^{-1}}) = 478 + (2.77\times 10^{-9}\times A_{\mathrm{CH}}(\mathrm{km^{2}}))\times (1-\frac{|\phi_{\mathrm{co}}(\mathrm{\degree})|}{61.4}),
\end{equation}

where $A_{\mathrm{CH}}$ denotes the CH area and $\phi_{\mathrm{co}}$ is the co-latitude of the center-of-mass of the CH and the planet projected onto the solar disk. We use that relation to calculate the SW speed fit, $v_{\mathrm{fit}}$ and compare that to the measured SW speed at 1 and 1.5 AU. The resulting scatter plot and correlation coefficients for Earth and Mars are given in Fig.~\ref{cc_area_v_fit}. The left panel shows the dependence of measured maximum bulk speed on CH area, resulting in a higher correlation coefficient for Mars ($cc$=0.49) than for Earth ($cc$=0.36). The differences between the measured maximum speed values and $v_{\mathrm{fit}}$ for Earth and Mars are shown in the right panel of Fig.~\ref{cc_area_v_fit}. These results may be applied in the future to set up a simple forecasting tool for SW HSSs at Mars based on Eq.~\ref{vfit}.

Figure~\ref{cc_width_SD} shows the dependence of the duration of the HSS on the longitudinal extent of the CH for Earth and Mars. For the CH longitudinal extension CH we derive a median of $26.5 \degree$, and except for one outlier, the widths are smaller than $50 \degree$. The median duration of the corresponding HSSs is obtained with $\sim$4.3 d at Earth and $\sim$4.1 d at Mars. The correlation between the two parameters is rather weak with $cc$=0.35 for Earth and $cc$=0.36 for Mars.

The relation between CH longitudinal extent and peak SW speed is given in the left panel of Fig.~\ref{cc_CH_speed}. Although a trend toward higher maximum speed is visible for CHs of higher longitudinal extent for both planets, the presence of a relatively large CH does not necessarily mean exceptionally high SW speed, for example, a CH of $15 \degree$ longitudinal extent may produce streams of similar speed as a CH of $40 \degree$ longitudinal extent. However, there seems to be a minimum threshold of SW peak speeds related to the CH extent, which is indicated in both panels of Fig.~\ref{cc_CH_speed} by a dashed line. The correlations are similar for Earth ($cc$=0.42) and Mars ($cc$=0.39).

In comparison, the right panel of Fig.~\ref{cc_CH_speed} gives the relation of the SW peak speed to the latitudinal extent of the CH. For both planets we derive a higher correlation between the SW peak speed with the latitudinal than with the longitudinal extent of the CH. Especially for CHs smaller than $<50 \degree$ in longitude, the speed of the HSS is stronger influenced by the North-South extent than by the East-West extent. The correlation is found to be weaker for Mars ($cc$=0.52) than for Earth ($cc$=0.64).
 
For each event during the opposition phases of 2016 and 2018, we extract the shock type from databases (CfA: online database of interplanetary shocks observed by the Wind and ACE spacecraft maintained by the Harvard-Smithsonian Center for Astrophysics \url{https://www.cfa.harvard.edu/shocks/} and IPS: Interplanetary shock database maintained by University of Helsinki \url{http://ipshocks.fi}). The result is summarized in Table ~\ref{SIRshocks} giving the occurrence rates of fast forward and fast reverse shocks at Earth and Mars. While the percentage of SIRs/CIRs connected with only fast reverse shocks is the same, the rate of forward shocks increases to a value of $\sim3$ times as much from Earth to Mars (i.e., from 6.7\% to 20.0\%). In the case of Earth, no events were observed that featured both a forward and a reverse shock, while at Mars there were even more events featuring both shocks than only a reverse shock (8.9\% vs 6.7\%).

\begin{table}
\caption{Shock type and occurrence rate for the 42 SIR events detected during the aligned phases of 2016 and 2018 at 1 and 1.5~AU extracted from CfA and IPS. \textit{FF} denotes fast forward shocks and \textit{FR} denotes fast reverse shocks. The absolute number of occurrences is given in parentheses next to the percentage.}
\label{SIRshocks}
\centering
\begin{tabular}{c c c}
\hline \hline
Shock type & Earth & Mars\\
\hline
FF only & 6.7\% (3)& 20.0\% (9)\\
FR only & 6.7\% (3) & 6.7\% (3)\\
FF and FR & 0 \% (0) & 8.9\% (4)\\\hline
FF and/or FR & 13.3\% (6) & 35.6\% (16)\\
\hline
\end{tabular}
\end{table}

\section{Discussion} \label{discussion}

We prepared a consistent dataset of 146 SIR/CIR events at Earth and 126 at Mars that occurred during the time range from November 2014 to November 2018 and investigate their evolution from 1 to 1.5~AU. We performed an SEA for the full sample of data and performed a more detailed analysis including the solar source information of CHs, for a subsample of 42 aligned events that occurred during the special opposition phases of the planets. 

\subsection{Superposed epoch analysis}

The SEA profiles show overall similar features for SW parameters at both planets. All absolute values show a clear drop from 1 to 1.5~AU. This is also obtained when taking only the aligned events into account - the magnitudes slightly drop, while the time differences between density peak and parameter peak are similar. \citet[][referred to as VB in the following]{Venzmer18} derived exponents for the  decrease/increase of SW parameters from 0.3--1AU (based on data from Helios 1+2 and OMNI). For the further distance up to 1.5AU, we find similar exponents for magnetic field strength $B_{avg} \varpropto r^{-1.63}$ (VB: $r^{-1.55}$) and number density $n_{avg} \varpropto r^{-1.99}$ (VB: $r^{-2.01}$). However, for the temperature, we obtain a larger decrease $T_{avg} \varpropto r^{-1.92}$ (VB: $r^{-0.79}$). Regarding the bulk speed, our results imply a decrease $v_{avg} \varpropto r^{-0.53}$ (VB: $r^{-0.05}$). This means that magnetic field strength and number density follow the modeled decrease in magnitude up to at least 1.52 AU, implying a merely geometrical evolution, also for SIRs/CIRs. In the case of temperature and bulk speed, the magnitudes at 1.52 AU are not simply a consequence of SW propagation through the heliosphere, but are also determined by SIR/CIR evolution effects. 

We further obtain that, on average, the speed profile of the stream is only weakly steepening from Earth to Mars given by the slightly shorter time difference from the velocity turning point to the maximum peak for Mars (cf.\,Table~\ref{SIRprops}). Pressure and magnetic field magnitude show, within the uncertainties, a rather simultaneous peak at Earth and Mars inferring the dominance of the magnetic pressure. The temperature maximum roughly co-moves with the maximum in the magnetic field magnitude and the time difference for the magnetic field peak at Mars is about $+$0.2 d compared to Earth. The maximum phase of the $P_{\mathrm{t}}$ (i.e., the SI) is broader for Mars than for Earth, hence, on average the SIR can be considered to be broader at Mars.

\begin{figure*}
\centering
\includegraphics[width = 1.\textwidth]{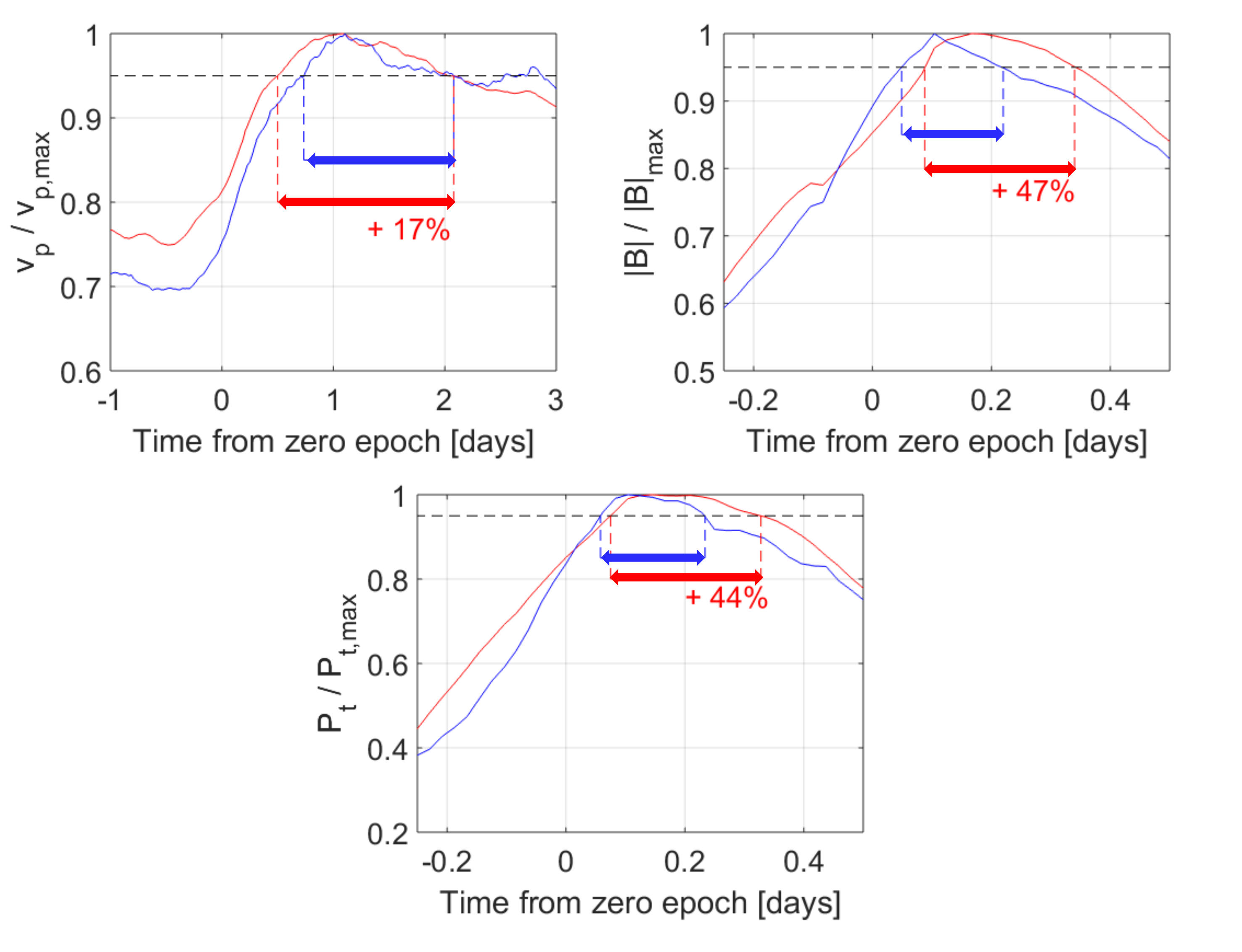}
\caption{Top left panel: Smoothed and normalized median of the speed profile at Earth (blue line) and Mars (red line). The black dashed line gives the threshold of 95\% of the maximum. The colored dashed lines show the respective duration of the curve above this threshold. The broadening of the wave crest for the median speed from Earth to Mars is then about 17\%. The epoch time ranges from -1 d to +3 d. Top right panel: Median of IMF magnitude around the stream interface. The marked features show an expansion of $\sim$47\%. The epoch time interval in this case is -0.25 d to +0.5 d. Bottom panel: Median profiles of the total perpendicular pressure, given again from -0.25 d to +0.5 d. The expansion rate for this parameter is about 43\%.}
\label{SEA_broad}
\end{figure*}

However, the overall duration of the stream is not revealing a broadening (4.3 d at Earth and 4.1 d at Mars as defined by the time that is needed until the SW speed drops below the average minimum speed; cf.\,Sect.~\ref{aligned_discuss}). \cite{Huang19} reported a SIR duration\footnote{Defined as the time range that lies between the pressure waves ahead and behind the SIR or, if present, between fast forward and reverse shocks.} that is similar for Earth ($\sim$36.7 h) and Mars ($\sim$37.0 h). In accordance to that, we determine from Fig.~\ref{vmed}, on average an SIR duration of $\sim$33.5 h in the case of Earth and $\sim$35.8 h in the case of Mars. On the other hand, we reveal a broadening of the HSS crest when comparing the profiles between 1 and 1.5
~AU. Figure~\ref{SEA_broad} shows the smoothed and normalized median profiles for speed, IMF magnitude and total perpendicular pressure. As we are interested in the crest part, we cut the profiles at a 95\% level from the maximum and compare their duration at Earth and Mars. For speed we derive an increase in duration by $\sim$17\%, and for the magnetic field and total perpendicular pressure an expansion of $\sim$47\% and $\sim$44\%, respectively. From that we conclude that the wave crest broadens related to the expansion of the stream close to the SI \citep[e.g.,][]{Gosling99}, which also explains the increased occurrence rate of fast-forward and also reverse shocks observed at Mars \citep[see also][]{Huang19}.

\begin{figure*}
\begin{subfigure}[b]{.5\textwidth}
\includegraphics[width=1.\textwidth]{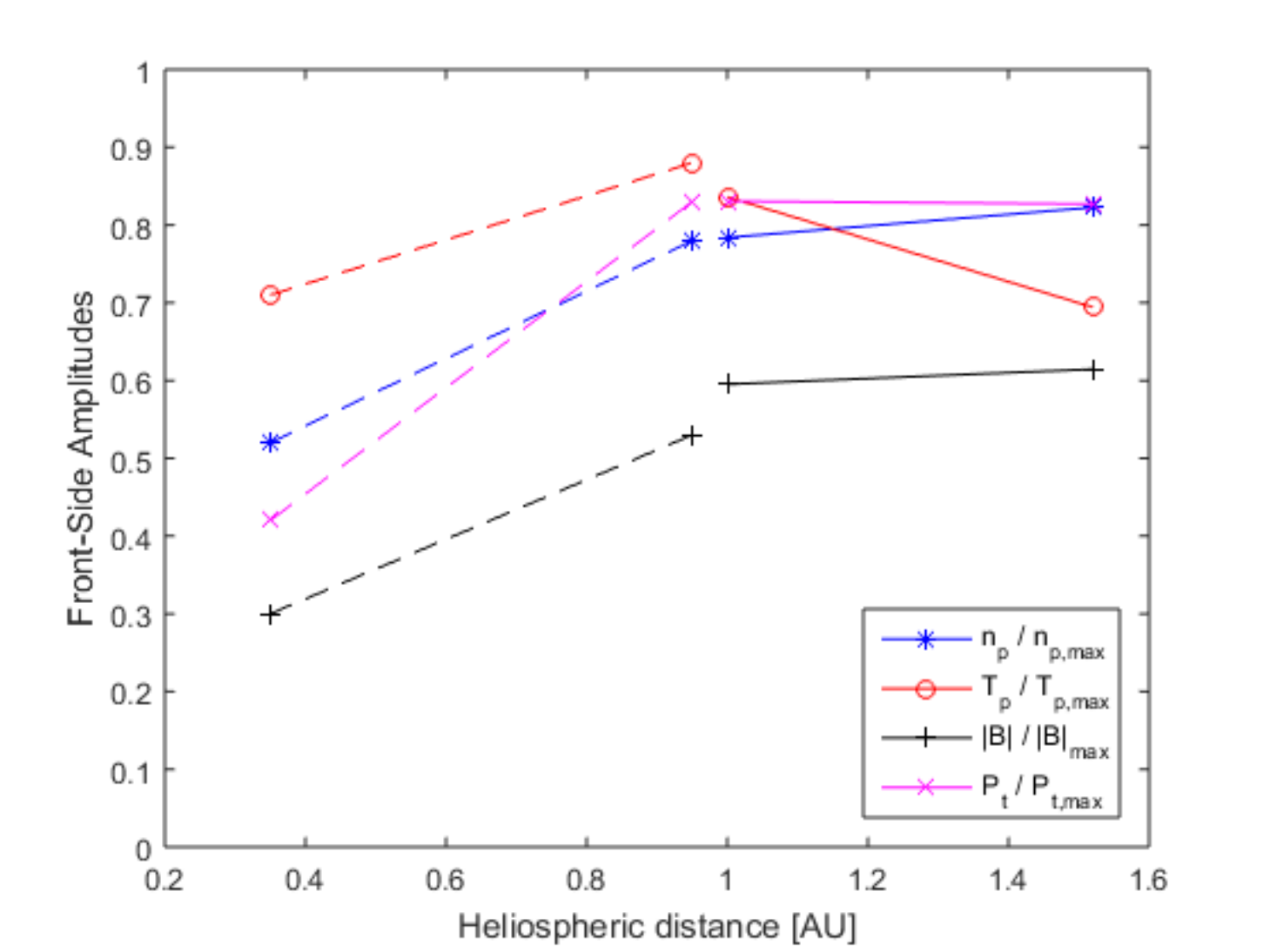}
\end{subfigure}%
\begin{subfigure}[b]{.5\textwidth}
\includegraphics[width=1.\textwidth]{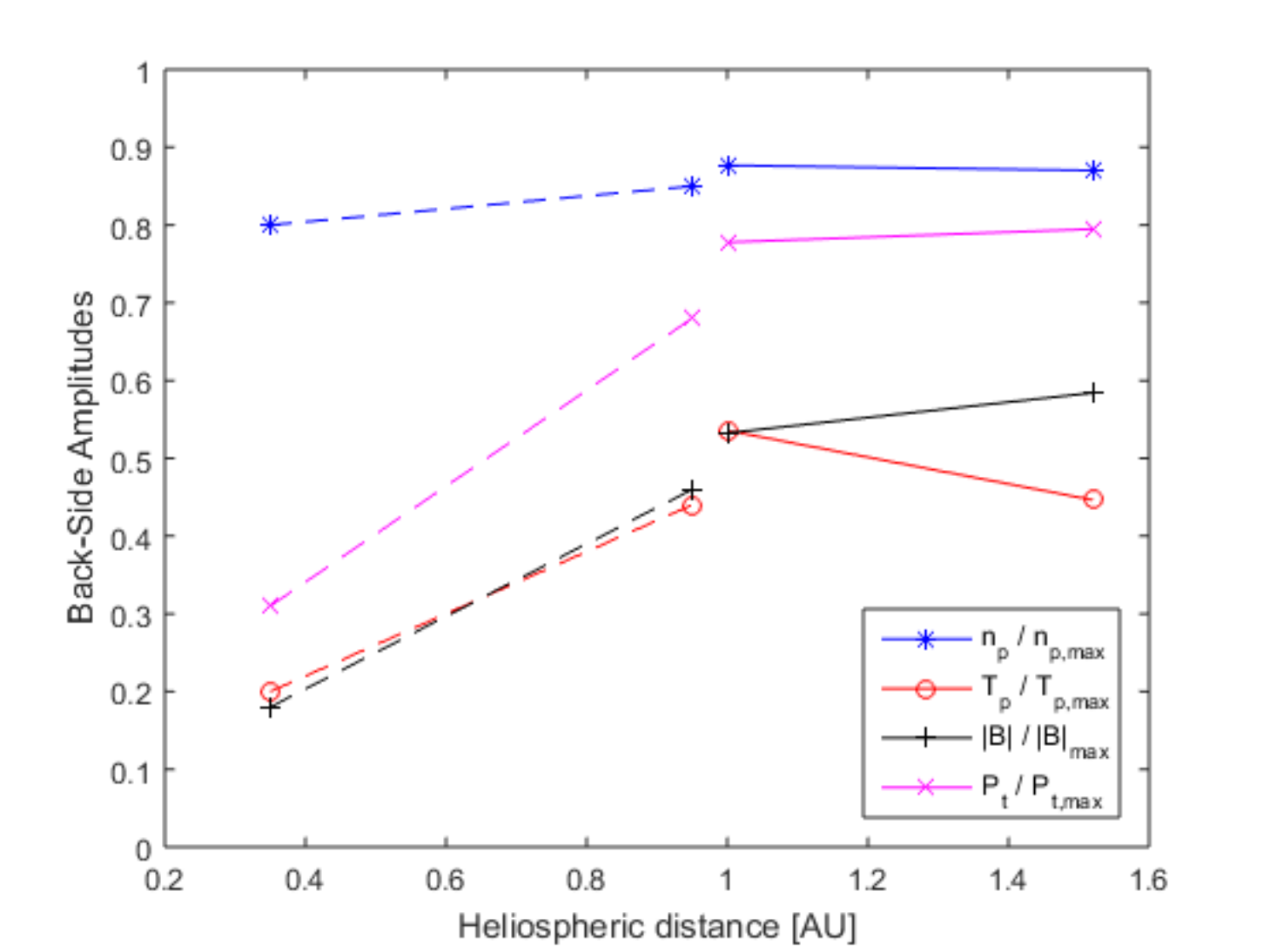}
\end{subfigure}
\caption{Amplitude (difference between minimum and maximum values) of the normalized SW parameters (given in the legend) taken across before zero epoch (front-side; left panel) and across after zero epoch (back-side; right panel) of the SEA profiles. Values for the distances 0.35 and 0.95~AU are taken from the study by \citet{richter86}.}
\label{richter2}
\end{figure*}

\cite{richter86} investigated Helios data (0.3--1.0~AU distance range) and found that the amplitudes of the SW parameters normalized to the maximum value all increase with increasing radial distance from the Sun. In analogy, we calculate the normalized amplitudes as derived from the presented SEA analysis for 1.0 and 1.5~AU and compare to the results given by \citet{richter86} for 0.35 and 0.95~AU (Fig.~\ref{richter2}). We obtain that the normalized amplitudes for density and magnetic field magnitude might continue the increasing trend as reported by \citet{richter86}, though with a much flatter slope. However, for the pressure component a clear stagnation is revealed and the temperature even strongly decreases beyond 1~AU. We conclude, beyond 1~AU the amplitudes do not follow a clear linear increase anymore but rather stagnate and plasma cools rapidly due to the further expansion of the stream. 

A striking feature of a strong drop in the IMF magnitude is observed at the time of zero epoch, identical to the time of the maximum median bulk density (Figs.~\ref{nmed} and \ref{Babsmed}). A similar feature has been reported for heliospheric plasma sheets located close to the HCS, hence, before the stream interface, and is explained by pressure balance, where a drop in magnetic field magnitude is connected with a rise in temperature and/or density \citep{Crooker04}. From the SEA for the total pressure we find a slight dip at the time of the magnetic field drop (Fig.~\ref{Ptmed}) together with a clear decrease in the entropy ($S=ln(T_p^{3/2}/N_p)$; not shown), meaning that pressure balancing is most likely not the only cause of the magnetic decrease observed in the region between the stream interface and the HSS. Other explanations for the magnetic drop could be magnetic holes or blob-like plasma structures, that might be formed close to the stream interface region \citep{Tsubouchi09,neugebauer04,jian19}. 

\subsection{Aligned events analysis}\label{aligned_discuss}

For the aligned events we studied 42 SIR/CIRs from Sun to Mars. \citet{Garton18} analyzed the dependence of SW speed measured near-Earth on the longitudinal extent of CHs during the years 2016 and 2017. In comparison with their results, the maximum stream duration of the HSS analyzed in this study is 8 d \citep[versus 14 in][]{Garton18}, while the longitudinal extent of the CHs does not exceed $80\ \degree$ (versus $120\ \degree$) for both planets. The linear correlation coefficient is weaker compared to \citet{Garton18} and nearly the same for both planets. However, we note that the correlation is strongly influenced by the most extended CH in longitude (but that is rather narrow in latitude), which produced a relatively medium peak SW speed at both planets. That CH also lies in the width regime ($>67\ \degree$ in longitude) for which a saturation of peak SW speed is found by \citet{Garton18} and supported by simulations as given in \citet{hofmeister20}. Interestingly, the correlation of the maximum bulk SW speed with the CH latitudinal extent is greater than that with the CH longitudinal extent for both planets. This implies a stronger dependence of the peak SW speed on the N-S extension for CHs with small to medium longitudinal extensions.

We find for the aligned events, having therefore weak effects of temporal evolution, a higher shock occurrence rate for Mars than for Earth (see Table~\ref{SIRshocks}) that is comparable to the results given in \citet{Huang19}. The percentage of streams associated with fast reverse shocks suggests no strong evolution at the trailing edge of SIRs at the same time. 
The aligned events also reveal a similar overall HSS duration between Earth and Mars (4.3 d at Earth and 4.1 d at Mars). We also derive at both planets similar correlations between stream and CH properties. We note, that the rather small longitudinal extent of the CHs in the studied time range leads in general to a shorter stream duration, hence, these results might be limited to HSSs emanating from small- to medium-size CHs. From the SEA we find that the expansion of the stream is most dominant around the SI as it evolves from Earth to Mars (see Table~\ref{SIRprops}). From that we conclude that the major changes in the stream evolution over radial distance occur at the front of SIRs close to the SI, leading to an increased rate of forward shocks. Merging processes of small SIRs into a strong SIR occurring over distance ranges of several AU may also lead to an increased shock rate of SIRs \citep{jian11}.

\cite{Lee17} give a general overview of Space Weather events at Mars during the period November 2014 -- May 2016 covering CMEs and CIRs in relation to flares and SEP events. They describe the recurrent CHs and CIR properties at Earth and Mars during the first conjunction period from May to July 2015. Here we show as complementary analysis a complete list of recurrent CHs during two opposition periods and their relation between Earth and Mars in-situ signatures (see Figs.~\ref{1_2016} to \ref{8_2018}). \citet{Lee17} analyzed a CIR observed by MAVEN over two rotations in June and July 2015. The source CH was found to be of medium longitudinal extension, similar to the CHs identified in the present study. As the CH declined in area, so did the bulk speed of the associated HSS, from $\sim$ 750 $\mathrm{km s^{-1}}$ to $\sim$ 522 $\mathrm{km s^{-1}}$. The density also decreased, from $\sim$ 38.5 $\mathrm{cm^{-3}}$ to $\sim$ 5.7 $\mathrm{cm^{-3}}$. This highly dynamical nature of CIRs can also be seen in Fig. \ref{CIR2}, as from the first to the second appearance, bulk speed and density are greatly reduced at both planets, independent of growing latitudinal separation. In general, SIR/CIR in-situ signatures look similar at both planets during opposition phases and also show similar evolution over one solar rotation. Also, the latitudinal separation is small ($<5\degree$ during the opposition phases) and the change of SIR 3D morphology might be at a larger latitudinal scale.

\section{Conclusions}
In conclusion, we find an expansion of the SIR/CIR most prominently visible in the total perpendicular pressure and the magnetic field strength. The amplitudes of SW parameters show a flatter increase from 1 to 1.52 AU than from 0.35 to 0.95 AU or even a decrease in the case of proton temperature. The decrease of peak number density and magnetic field strength from 1 to 1.52 AU scales according to a geometrical expansion, while proton temperature and speed show additional evolution effects. The dependence of stream duration and bulk speed on the properties of the source CH is similar for both planets, and reveal moderate correlation coefficients between stream and CH properties. The fast forward shock occurrence rate increases by a factor of 3 from Earth to Mars, while no instance of SIRs featuring both fast forward and fast reverse shocks have been observed at Earth during the aligned events period. The findings of this study can be used in the future to not only further expand the knowledge about the evolution of SIRs and CIRs beyond 1 AU, but also serve for modeling of SIR/CIR propagation from the Sun to the outer heliosphere. Studying the evolution of SW streams throughout the heliosphere is essential for better understanding the large-scale heliospheric magnetic structure and the space weather conditions at various planets where future space explorations will take place.

\begin{acknowledgements} We thank Lan Jian for very valuable discussions. J.G. thanks Jasper Halekas for the support of MAVEN data. 
We also thank the reviewer for the careful revision of this manuscript to improve its content. J.G. is supported by the Strategic Priority Program of the Chinese Academy of Sciences (Grant No. XDB41000000), the National Natural Science Foundation of China (Grant No. 42074222) and the CNSA pre-research Project on Civil Aerospace Technologies (Grant No. D020104).
The MAVEN data are archived in the NASA Planetary Data System’s Planetary Plasma Interactions Node hosted at https://pds-ppi.igpp.ucla.edu/.
\end{acknowledgements}

\bibliographystyle{aa}

\begin{appendix}
\section{Aligned events}
Here we give more detailed information about the aligned events during the opposition phase of Mars and Earth in 2016 and 2018, respectively. We list in the following additional Tables for exact dates and occurrence frequency of the SIRs and CIRs. The Figures show corresponding EUV image data and in-situ measured solar wind speed and density profiles for each CIR. 

\begin{table*}
  \centering
  \caption{Catalog of aligned SIRs and CIRs for 2016. Note. The units are as follows: $N_{\mathrm{max}}\  [\mathrm{cm^-{3}}]$,  $v_{\mathrm{max}}\  [\mathrm{km\ s^{-1}}]$,  $B_{\mathrm{max}}\  [\mathrm{nT}]$,  $T_{\mathrm{max}}\ [10^{4}\ \mathrm{K}]$,  $P_{\mathrm{t,max}}\  [\mathrm{pPa}]$,  $\Phi_{\mathrm{HG}}$ and $\Theta_{\mathrm{HG}} [\degree] $, and denote the maximum of plasma density, bulk speed, IMF magnitude, proton temperature and total perpendicular pressure, respectively. The two last columns denote heliographic latitude and longitude of the planets. The occurrence of a fast forward (\textit{FF}) or reverse (\textit{FR}) shock is indicated with a Y. Shocks were allocated using CfA and IPS databases (Earth) and \citet{Huang19} (Mars).}
    \begin{tabular}{rrrrrrrrrrrrrr}
          \multicolumn{1}{l}{Planet} & \multicolumn{1}{l}{SIR \#} & \multicolumn{1}{l}{CIR \#} & \multicolumn{1}{l}{HSS Start} & \multicolumn{1}{l}{HSS End} &
          \multicolumn{1}{l}{FF?} &
          \multicolumn{1}{l}{FR?} &
          \multicolumn{1}{l}{$N_{\mathrm{max}}$} & \multicolumn{1}{l}{$V_{\mathrm{max}}$} & \multicolumn{1}{l}{$B_{\mathrm{max}}$} & \multicolumn{1}{l}{$T_{\mathrm{max}}$} & \multicolumn{1}{l}{$P_{\mathrm{t,max}}$} & \multicolumn{1}{l}{$\Phi_{\mathrm{HG}}$} & \multicolumn{1}{l}{$\Theta_{\mathrm{HG}}$} \\
          \hline
    \multicolumn{1}{l}{Earth} & 1     &       & 03-06 18:20 & 03-09 22:17 &       &       & 33.4  & 632.7 & 20.9  & 85.2  & 359.5 & -7.2  & 138.3 \\
    \multicolumn{1}{l}{Mars} &       &       & 03-11 11:53 & 03-15 13:55 & Y     &       & 17.2  & 588.6 & 9.7   & 32.0  & 58.7  & -4.9  & 176.3 \\
    \hdashline
          & 2     &       & 03-11 11:21 & 03-14 14:07 &       & Y     & 71.9  & 591.8 & 26.2  & 49.7  & 328.1 & -7.2  & 72.5 \\
          &       &       & 03-16 12:50 & 03-19 03:33 &       &       & 3.6   & 513.2 & 7.4   & 23.5  & 25.6  & -4.8  & 107.7 \\
          \hdashline
          & 3     &       & 03-14 18:34 & 03-22 01:49 & Y     &       & 70.2  & 629.8 & 24.3  & 108.3 & 347.8 & -7.2  & 32.9 \\
          &       &       & 03-20 01:23 & 03-26 15:05 &       &       & 10.7  & 577.6 & 7.9   & 29.4  & 37.7  & -4.7  & 66.6 \\\hdashline
          & 4     & 1     & 05-27 00:31 & 06-02 21:00 &       &       & 27.2  & 577.6 & 13.1  & 32.9  & 73.5  & -1.2  & 135.6 \\
          &       &       & 05-29 02:40 & 06-05 01:09 &       &       & 12.4  & 471.8 & 5.5   & 20.0  & 26.1  & -1.9  & 133.7 \\\hdashline
          & 5     & 2     & 06-05 09:09 & 06-09 20:42 &       &       & 71.4  & 674.6 & 21.2  & 118.2 & 310.1 & -0.1  & 16.5 \\
          &       &       & 06-07 01:39 & 06-11 17:06 & Y     & Y     & 39.2  & 550.1 & 14.6  & 33.9  & 96.3  & -1.5  & 10.8 \\\hdashline
          & 6     & 3     & 06-10 13:26 & 06-14 10:23 &       &       & 27.0  & 574.8 & 14.0  & 83.0  & 125.8 & 0.5   & 310.3 \\
          &       &       & 06-12 12:11 & 06-15 08:39 & Y     &       & 15.6  & 483.5 & 7.8   & 22.4  & 30.9  & -1.2  & 302.5 \\\hdashline
          & 7     &       & 06-14 19:49 & 06-22 02:42 & Y     &       & 25.7  & 728.2 & 16.5  & 50.6  & 133.0 & 0.9   & 257.4 \\
          &       &       & 06-15 08:54 & 06-22 09:14 &       &       & 4.2   & 607.3 & 5.8   & 38.1  & 23.3  & -1.0  & 248.0 \\\hdashline
          & 8     & 1     & 06-22 20:11 & 06-23 06:21 &       &       & 43.0  & 440.2 & 14.1  & 37.2  & 96.6  & 1.9   & 151.5 \\
          &       &       & 06-23 08:47 & 06-24 05:25 & Y     &       & 8.0   & 424.0 & 5.3   & 12.7  & 14.9  & -0.6  & 138.8 \\\hdashline
          & 9     &       & 06-23 08:13 & 06-29 21:14 &       &       & 15.3  & 564.2 & 10.6  & 42.4  & 55.5  & 2.0   & 138.2 \\
          &       &       & 06-24 10:25 & 06-30 10:51 &       &       & 5.6   & 504.6 & 7.9   & 24.7  & 28.0  & -0.5  & 125.2 \\\hdashline
          & 10    & 2     & 07-02 22:40 & 07-05 15:10 &       &       & 51.7  & 507.9 & 15.8  & 38.9  & 149.3 & 3.0   & 19.1 \\
          &       &       & 07-03 11:37 & 07-06 22:12 &       & Y     & 19.3  & 498.8 & 8.3   & 23.9  & 44.9  & 0.0   & 2.5 \\\hdashline
          & 11    & 3     & 07-07 01:17 & 07-11 22:01 &       &       & 49.0  & 659.7 & 14.2  & 42.3  & 120.1 & 3.5   & 312.9 \\
          &       &       & 07-07 19:21 & 07-11 07:27 & Y     &       & 18.3  & 653.0 & 13.8  & 34.9  & 106.9 & 0.2   & 294.4 \\\hdashline
          & 12    &       & 07-12 05:59 & 07-14 02:48 &       &       & 6.3   & 659.4 & 8.8   & 37.3  & 40.5  & 4.1   & 246.8 \\
          &       &       & 07-13 03:56 & 07-13 23:25 &       &       & 4.2   & 580.1 & 5.8   & 24.6  & 16.4  & 0.5   & 226.3 \\\hdashline
          & 13    & 4     & 07-14 13:55 & 07-18 23:12 &       &       & 5.0   & 713.3 & 7.4   & 31.4  & 31.8  & 4.3   & 220.3 \\
          &       &       & 07-14 11:10 & 07-18 23:21 &       &       & 3.6   & 605.3 & 6.4   & 24.1  & 25.5  & 0.6   & 199.1 \\\hdashline
          & 14    & 2     & 07-28 04:22 & 08-01 04:25 &       &       & 50.6  & 646.4 & 17.7  & 88.0  & 236.3 & 5.5   & 35.1 \\
          &       &       & 07-28 18:21 & 08-01 02:00 &       &       & 15.7  & 528.6 & 9.0   & 22.9  & 36.9  & 1.4   & 8.6 \\\hdashline
          & 15    & 4     & 08-09 03:47 & 08-15 08:52 &       &       & 7.2   & 673.1 & 8.6   & 41.4  & 40.2  & 6.3   & 236.4 \\
          &       &       & 08-08 15:32 & 08-14 08:08 &       &       & 3.0   & 641.4 & 4.9   & 29.8  & 16.0  & 2.1   & 205.4 \\\hdashline
          & 16    & 1     & 08-17 05:02 & 08-18 09:51 &       &       & 39.1  & 444.5 & 12.3  & 16.4  & 69.7  & 6.7   & 130.6 \\
          &       &       & 08-15 22:08 & 08-20 00:34 &       &       & 7.8   & 394.1 & 5.8   & 12.5  & 14.8  & 2.5   & 96.7 \\\hdashline
          & 17    & 2     & 08-23 18:45 & 08-29 03:49 &       &       & 18.0  & 606.7 & 16.2  & 51.9  & 146.5 & 7.0   & 51.3 \\
          &       &       & 08-22 14:27 & 08-27 00:14 &       &       & 9.0   & 531.2 & 6.8   & 26.9  & 28.4  & 2.8   & 15.2 \\\hdashline
          & 18    & 3     & 08-29 11:00 & 08-31 20:14 &       &       & 29.1  & 483.8 & 10.0  & 18.7  & 49.2  & 7.1   & 332.0 \\
          &       &       & 08-27 05:05 & 08-30 12:14 & Y     &       & 12.8  & 543.6 & 7.1   & 22.6  & 27.7  & 3.1   & 293.8 \\\hdashline
          & 19    &       & 09-01 05:23 & 09-10 17:10 &       &       & 12.4  & 742.0 & 11.1  & 49.3  & 84.6  & 7.2   & 305.6 \\
          &       &       & 08-30 14:51 & 09-07 10:20 &       &       & 5.9   & 576.7 & 6.4   & 27.1  & 25.4  & 3.2   & 266.6 \\\hdashline
          & 20    & 2     & 09-19 21:11 & 09-24 10:26 &       &       & 38.4  & 773.9 & 21.9  & 110.4 & 252.9 & 7.1   & 41.5 \\
          &       &       & 09-17 23:21 & 09-22 23:46 & Y     & Y     & 31.7  & 698.3 & 13.3  & 34.8  & 97.6  & 4.1   & 355.3 \\
\end{tabular}%
  \label{AEA_table2016}%
\end{table*}%

\begin{table*}
  \centering
  \caption{Catalog of aligned SIRs and CIRs for 2018. Note. The units are as follows: $N_{\mathrm{max}}\  [\mathrm{cm^-{3}}]$,  $v_{\mathrm{max}}\  [\mathrm{km\ s^{-1}}]$,  $B_{\mathrm{max}}\  [\mathrm{nT}]$,  $T_{\mathrm{max}}\ [10^{4}\ \mathrm{K}]$,  $P_{\mathrm{t,max}}\  [\mathrm{pPa}]$,  $\Phi_{\mathrm{HG}}$ and $\Theta_{\mathrm{HG}} [\degree] $. The occurrence of a fast forward (\textit{FF}) or reverse (\textit{FR}) shock is indicated with a Y. \textit{NU} denotes no upstream period for MAVEN.}
    \begin{tabular}{rrrrrrrrrrrrrrr}
          \multicolumn{1}{l}{Planet} & \multicolumn{1}{l}{SIR \#} & \multicolumn{1}{l}{CIR \#} & \multicolumn{1}{l}{HSS Start} & \multicolumn{1}{l}{HSS End} &
          \multicolumn{1}{l}{FF?} &
          \multicolumn{1}{l}{FR?} & \multicolumn{1}{l}{$N_{\mathrm{max}}$} & \multicolumn{1}{l}{$V_{\mathrm{max}}$} & \multicolumn{1}{l}{$B_{\mathrm{max}}$} & \multicolumn{1}{l}{$T_{\mathrm{max}}$} & \multicolumn{1}{l}{$P_{\mathrm{t,max}}$} & \multicolumn{1}{l}{$\Phi_{\mathrm{HG}}$} & \multicolumn{1}{l}{$\Theta_{\mathrm{HG}}$} \\
          \hline
    \multicolumn{1}{l}{Earth} & 21    &       & 03-29 15:10 & 04-03 15:22 &       &       & 28.1  & 469.0 & 8.7   & 21.7  & 46.5  & -6.7  & 280.0 \\
    \multicolumn{1}{l}{Mars} &       &       & 04-05 11:05 & 04-09 13:08 &       &       & 15.2  & 500.4 & 7.1   & 20.2  & 29.7  & -2.7  & 327.7 \\ \hdashline
          & 22    &       & 04-08 18:20 & 04-15 00:18 &       &       & 15.7  & 617.3 & 10.0  & 33.5  & 52.6  & -6.1  & 148.1 \\
          &       &       & 04-15 21:49 & 04-21 07:30 &       &       & 19.1  & 544.2 & 7.6   & 21.6  & 35.6  & -2.2  & 191.0 \\\hdashline
          & 23    &       & 04-20 04:23 & 04-24 10:48 & \multicolumn{1}{l}{Y} &       & 68.2  & 669.6 & 23.4  & 64.3  & 272.6 & -5.2  & 349.7 \\
          &       &       & 04-25 04:43 & \textit{NU} &       &       & 4.7   & 544.0 & 5.2   & 31.7  & 30.6  & -1.6  & 27.1 \\\hdashline
          & 24    & 5     & 06-26 08:05 & 06-26 19:58 &       &       & 25.3  & 636.3 & 14.7  & 45.4  & 103.0 & 2.3   & 183.4 \\
          &       &       & 06-27 20:04 & 06-28 22:54 &       &       & 7.2   & 601.6 & 5.1   & 26.6  & 15.1  & 2.0   & 194.4 \\\hdashline
          & 25    & 6     & 07-20 21:32 & 07-23 21:38 &       &       & 13.6  & 570.2 & 10.2  & 46.0  & 76.2  & 4.8   & 225.8 \\
          &       &       & 07-21 10:20 & 07-25 02:58 &       & \multicolumn{1}{l}{Y} & 6.1   & 507.4 & 6.7   & 27.0  & 29.5  & 3.3   & 228.4 \\\hdashline
          & 26    & 5     & 07-24 05:50 & 07-28 09:21 &       &       & 46.4  & 612.8 & 13.7  & 43.1  & 97.7  & 5.1   & 172.9 \\
          &       &       & 07-25 18:46 & 07-30 06:35 &       &       & 25.9  & 587.1 & 10.6  & 31.7  & 54.9  & 3.5   & 174.1 \\\hdashline
          & 27    & 7     & 07-31 19:27 & 08-02 01:27 &       &       & 24.3  & 422.3 & 9.8   & 26.1  & 43.0  & 5.7   & 80.3 \\
          &       &       & 08-01 16:42 & 08-03 04:00 &       &       & 7.8   & 332.9 & 5.6   & 10.5  & 13.9  & 3.8   & 79.2 \\\hdashline
          & 28    &       & 08-02 02:14 & 08-02 22:09 &       &       & 32.2  & 404.7 & 8.9   & 15.7  & 39.8  & 5.9   & 40.6 \\
          &       &       & 08-03 12:35 & 08-04 11:35 & \multicolumn{1}{l}{Y} &       & 11.3  & 362.7 & 6.7   & 13.2  & 20.5  & 3.9   & 38.5 \\\hdashline
          & 29    &       & 08-05 10:31 & 08-06 21:52 &       &       & 15.1  & 400.0 & 6.8   & 12.3  & 22.5  & 6.0   & 14.2 \\
          &       &       & 08-06 19:11 & 08-08 01:50 &       & \multicolumn{1}{l}{Y} & 11.8  & 424.6 & 6.3   & 16.2  & 20.2  & 4.0   & 11.3 \\\hdashline
          & 30    &       & 08-07 16:51 & 08-08 21:26 &       &       & 24.8  & 433.8 & 10.0  & 16.3  & 41.9  & 6.2   & 347.7 \\
          &       &       & 08-09 05:12 & 08-11 03:06 &       &       & 14.2  & 363.2 & 5.5   & 7.0   & 16.0  & 4.1   & 344.2 \\\hdashline
          & 31    & 8     & 08-11 07:37 & 08-14 01:32 &       &       & 35.6  & 425.3 & 11.4  & 15.4  & 60.9  & 6.5   & 281.6 \\
          &       &       & 08-12 04:49 & 08-14 23:51 & \multicolumn{1}{l}{Y} &       & 10.4  & 414.1 & 7.4   & 18.1  & 28.4  & 4.3   & 276.4 \\\hdashline
          & 32    & 6     & 08-14 23:59 & 08-19 15:13 &       &       & 26.3  & 575.0 & 11.8  & 26.0  & 71.9  & 6.6   & 255.2 \\
          &       &       & 08-16 07:08 & 08-20 03:44 &       &       & 17.8  & 540.2 & 7.8   & 23.9  & 37.3  & 4.4   & 249.3 \\\hdashline
          & 33    & 5     & 08-19 17:51 & 08-24 01:05 &       & \multicolumn{1}{l}{Y} & 25.0  & 672.8 & 12.9  & 58.9  & 97.4  & 6.8   & 189.1 \\
          &       &       & 08-20 14:57 & 08-25 19:11 &       &       & 13.6  & 646.2 & 11.0  & 39.1  & 60.8  & 4.6   & 181.5 \\\hdashline
          & 34    & 7     & 08-26 13:18 & 09-02 04:12 &       &       & 34.0  & 633.9 & 19.2  & 46.4  & 162.5 & 7.1   & 83.4 \\
          &       &       & 08-27 23:40 & 08-31 22:42 & \multicolumn{1}{l}{Y} &       & 24.9  & 544.2 & 10.8  & 23.2  & 59.6  & 4.8   & 73.1 \\\hdashline
          & 35    & 8     & 09-07 04:06 & 09-09 17:45 &       &       & 28.7  & 500.6 & 11.7  & 40.8  & 82.4  & 7.2   & 298.1 \\
          &       &       & 09-07 16:25 & 09-10 12:14 &       &       & 8.8   & 531.5 & 6.9   & 23.0  & 24.4  & 5.2   & 284.1 \\\hdashline
          & 36    & 6     & 09-10 16:05 & 09-16 20:55 &       &       & 26.2  & 703.5 & 15.4  & 51.1  & 119.2 & 7.2   & 258.4 \\
          &       &       & 09-11 17:53 & 09-17 02:59 & \multicolumn{1}{l}{Y} &       & 15.9  & 547.8 & 9.4   & 24.9  & 47.7  & 5.2   & 243.4 \\\hdashline
          & 37    & 5     & 09-17 02:26 & 09-20 15:30 &       &       & 37.0  & 585.6 & 11.9  & 43.5  & 85.5  & 7.2   & 166.0 \\
          &       &       & 09-17 11:02 & 09-21 16:18 &       &       & 17.1  & 526.1 & 8.6   & 27.8  & 34.8  & 5.4   & 148.6 \\\hdashline
          & 38    & 8     & 10-03 15:29 & 10-06 23:57 &       &       & 16.0  & 515.9 & 9.3   & 30.5  & 50.8  & 6.6   & 314.8 \\
          &       &       & 10-03 18:26 & 10-06 18:59 &       &       & 16.1  & 532.2 & 11.2  & 28.5  & 51.2  & 5.6   & 291.8 \\\hdashline
          & 39    & 6     & 10-07 10:45 & 10-12 22:04 &       &       & 66.9  & 640.1 & 18.2  & 44.2  & 161.4 & 6.4   & 262.1 \\
          &       &       & 10-07 02:01 & 10-11 17:19 &       &       & 12.6  & 601.2 & 8.1   & 28.9  & 46.3  & 5.6   & 237.6 \\\hdashline
          & 40    & 5     & 10-13 17:49 & 10-18 16:58 &       &       & 32.1  & 641.1 & 12.5  & 39.6  & 71.6  & 6.0   & 182.9 \\
          &       &       & 10-11 21:46 & 10-18 03:28 &       &       & 7.0   & 577.6 & 5.9   & 24.6  & 22.3  & 5.6   & 156.3 \\\hdashline
          & 41    & 6     & 11-04 21:01 & 11-09 15:41 &       &       & 24.4  & 638.5 & 13.6  & 54.9  & 152.5 & 4.1   & 252.8 \\
          &       &       & 11-03 14:58 & 11-08 00:13 & \multicolumn{1}{l}{Y} & \multicolumn{1}{l}{Y} & 15.8  & 541.7 & 8.7   & 22.7  & 40.2  & 5.5   & 218.1 \\\hdashline
          & 42    & 5     & 11-09 16:41 & 11-16 11:21 &       &       & 20.3  & 656.0 & 14.1  & 42.7  & 91.2  & 3.7   & 200.0 \\
          &       &       & 11-08 14:11 & 11-14 10:03 &       &       & 8.0   & 567.4 & 6.6   & 32.3  & 30.3  & 5.4   & 163.9 \\
    \end{tabular}%
  \label{AEA_table2018}
\end{table*}%

\begin{table*}
  \centering
  \caption{SIRs/CIRs ocurring at times of CME disturbances at Earth. Note. The units are as follows: $N_{\mathrm{max}}\  [\mathrm{cm^-{3}}]$,  $v_{\mathrm{max}}\  [\mathrm{km\ s^{-1}}]$,  $B_{\mathrm{max}}\  [\mathrm{nT}]$,  $T_{\mathrm{max}}\ [10^{4}\ \mathrm{K}]$,  $P_{\mathrm{t,max}}\  [\mathrm{pPa}]$,  $\Phi_{\mathrm{HG}}$ and $\Theta_{\mathrm{HG}} [\degree] $. The occurrence of a fast forward (\textit{FF}) or reverse (\textit{FR}) shock is indicated with a Y.}
    \begin{tabular}{rrrrrrrrrrrrr}
              \multicolumn{1}{l}{Planet} & \multicolumn{1}{l}{CIR \#} & \multicolumn{1}{l}{HSS Start} & \multicolumn{1}{l}{HSS End} &
          \multicolumn{1}{l}{FF?} &
          \multicolumn{1}{l}{FR?} & \multicolumn{1}{l}{$N_{\mathrm{max}}$} & \multicolumn{1}{l}{$V_{\mathrm{max}}$} & \multicolumn{1}{l}{$B_{\mathrm{max}}$} & \multicolumn{1}{l}{$T_{\mathrm{max}}$} & \multicolumn{1}{l}{$P_{\mathrm{t,max}}$} & \multicolumn{1}{l}{$\Phi_{\mathrm{HG}}$} & \multicolumn{1}{l}{$\Theta_{\mathrm{HG}}$} \\
          \hline
    \multicolumn{1}{l}{Earth} & 1     & 2016 07-20 12:36 & 07-24 07:26 &       &       & 70.5  & 603.4 & 34.3  & 85.1  & 555.8 & 4.8   & 140.9 \\
      \multicolumn{1}{l}{Mars}  &       & 2016 07-21 06:57 & 07-25 13:27 &       &       & 13.7  & 686.8 & 11.8  & 36.2  & 73.5  & 1.0   & 117.4 \\\hdashline
     &3     & 2016 08-02 11:34 & 08-08 08:31 &       & \multicolumn{1}{l}{Y} & 55.7  & 713.1 & 25.2  & 67.1  & 333.7 & 5.9   & 328.9 \\
          &       & 2016 08-02 04:05 & 08-08 09:02 & \multicolumn{1}{l}{Y} & \multicolumn{1}{l}{Y} & 2.5   & 707.0 & 6.4   & 6.5   & 18.5  & 1.7   & 300.6 \\\hdashline
    &7     & 2018 09-21 23:11 & 09-24 12:02 &       &       & 32.5  & 587.5 & 10.7  & 43.4  & 56.6  & 7.1   & 113.2 \\
          &       & 2018 09-22 10:47 & 09-25 05:00 &       &       & 12.5  & 566.5 & 10.2  & 32.4  & 46.9  & 5.4   & 94.4 \\
    \end{tabular}%
  \label{CME_table}%
\end{table*}%

\begin{figure*}
\centering
\includegraphics[width = \textwidth]{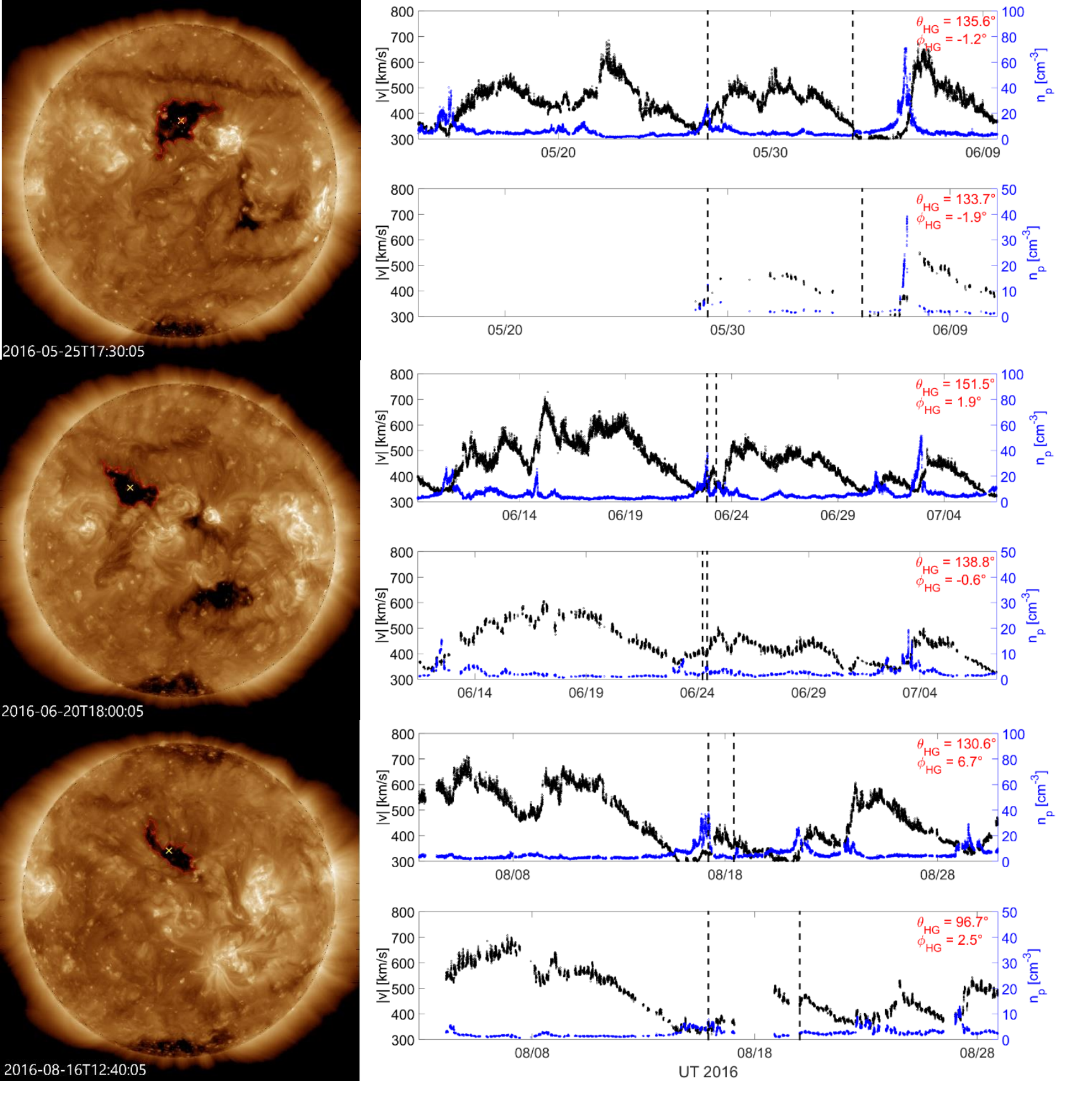}
\caption{The first corotating interaction region (CIR 1) that occurred during the opposition phase in 2016. \textit{Left panels}: SDO/AIA 193~\AA~images of the source CH extracted using CATCH. The via thresholding determined CH boundary is given in red, a yellow x marks the geometrical center-of-mass. \textit{Right panels}: In-situ SW data showing proton number density (blue) and bulk speed (black) for one synodic rotation. The odd panels show SW data obtained from the OMNI database, the even panels depict MAVEN data. The heliographic coordinates of the planets at the time of SI passage are given in red in the upper right corner.}
\label{1_2016}
\end{figure*}

\begin{figure*}
\centering
\includegraphics[width = \textwidth]{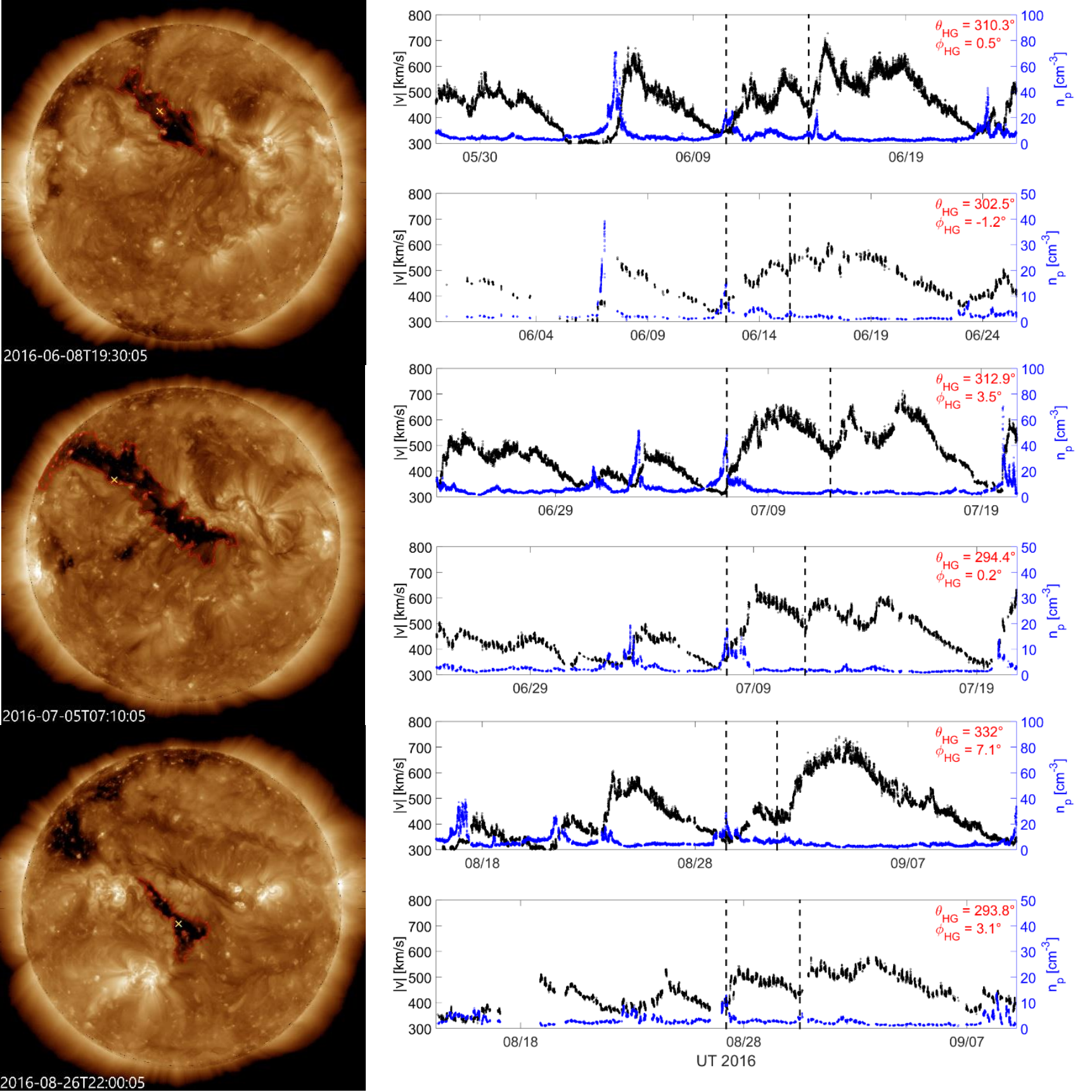}
\caption{Same as Fig.~\ref{1_2016} for CIR 3, occurring in 2016.}
\label{3_2016}
\end{figure*}

\begin{figure*}
\centering
\includegraphics[width = \textwidth]{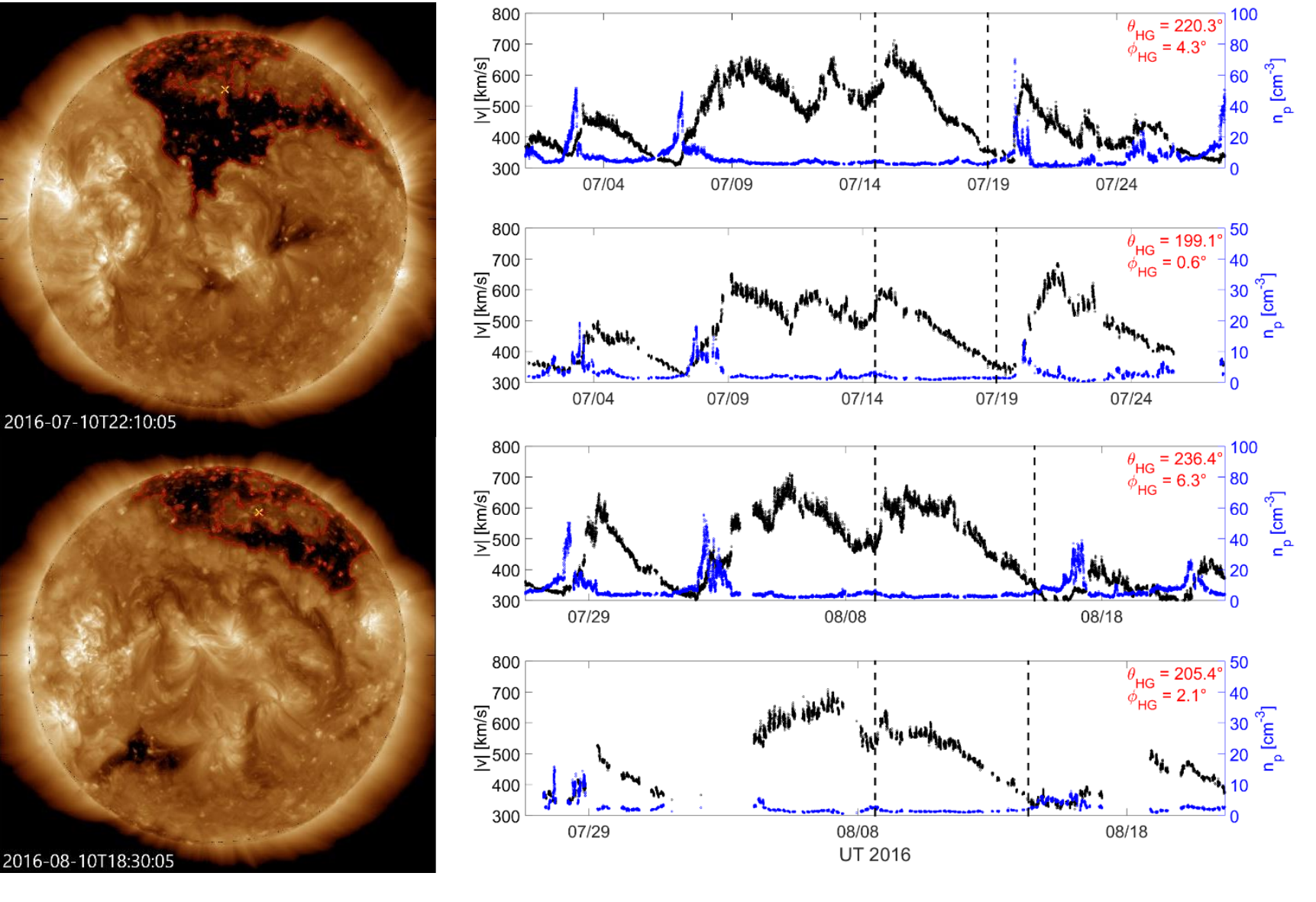}
\caption{Same as Fig.~\ref{1_2016} for CIR 4, occurring in 2016.}
\label{4_2016}
\end{figure*}

\begin{figure*}
\centering
\includegraphics[width = 0.7\textwidth]{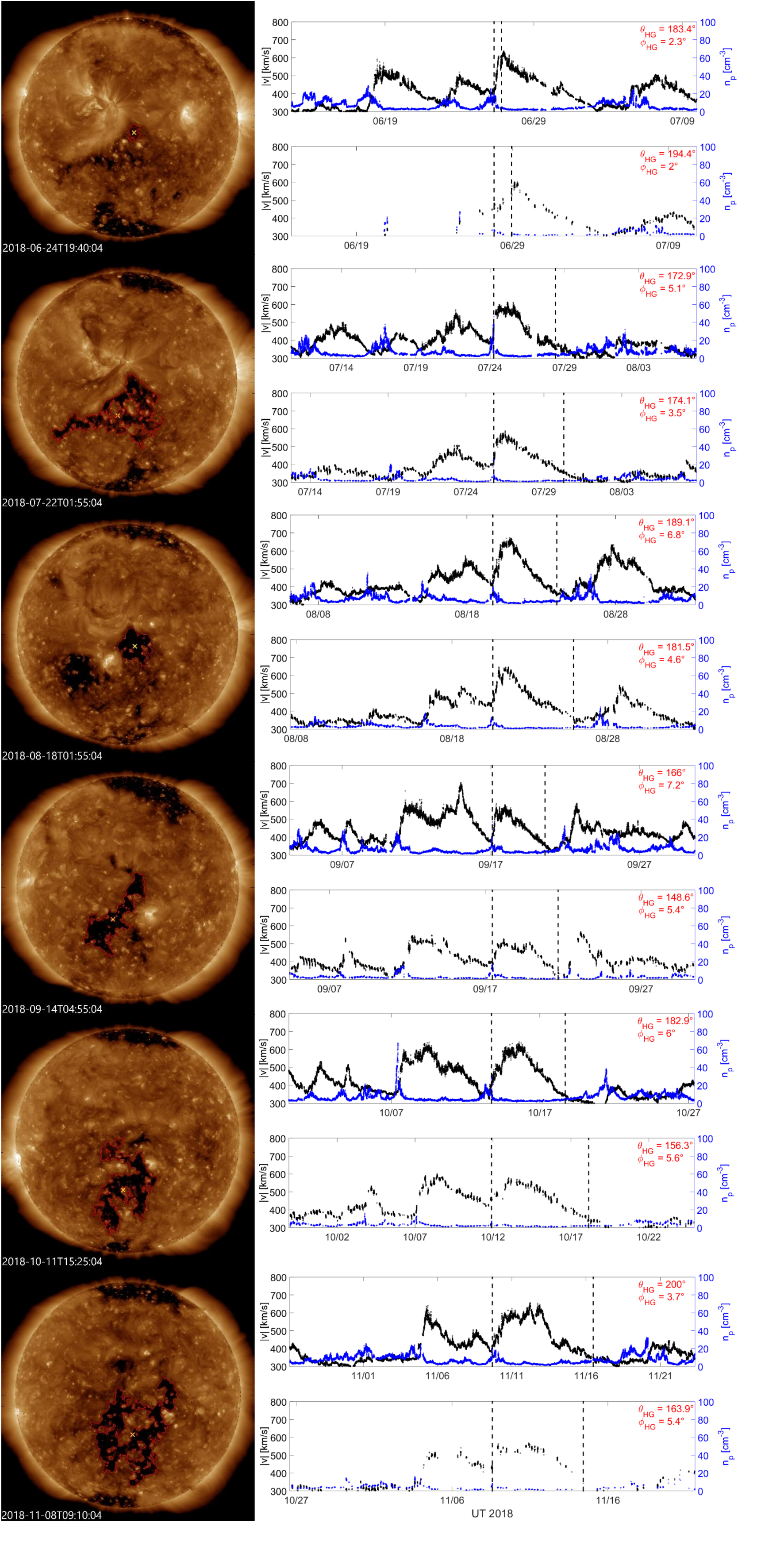}
\caption{Same as Fig.~\ref{1_2016} for CIR 5, occurring in 2018.}
\label{5_2018}
\end{figure*}

\begin{figure*}
\centering
\includegraphics[width = 0.8\textwidth]{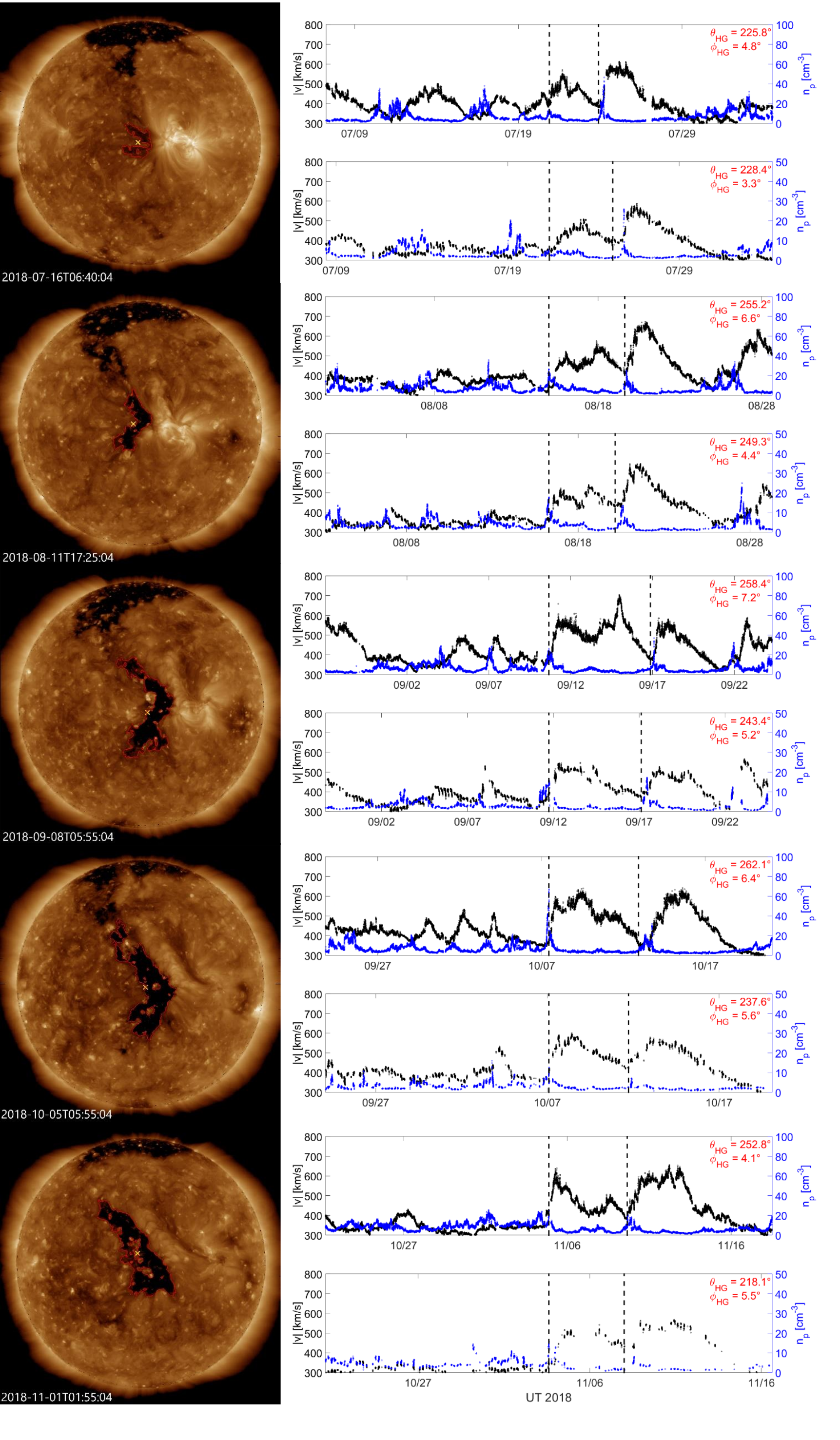}
\caption{Same as Fig.~\ref{1_2016} for CIR 6, occurring in 2018.}
\label{6_2018}
\end{figure*}

\begin{figure*}
\centering
\includegraphics[width =  \textwidth]{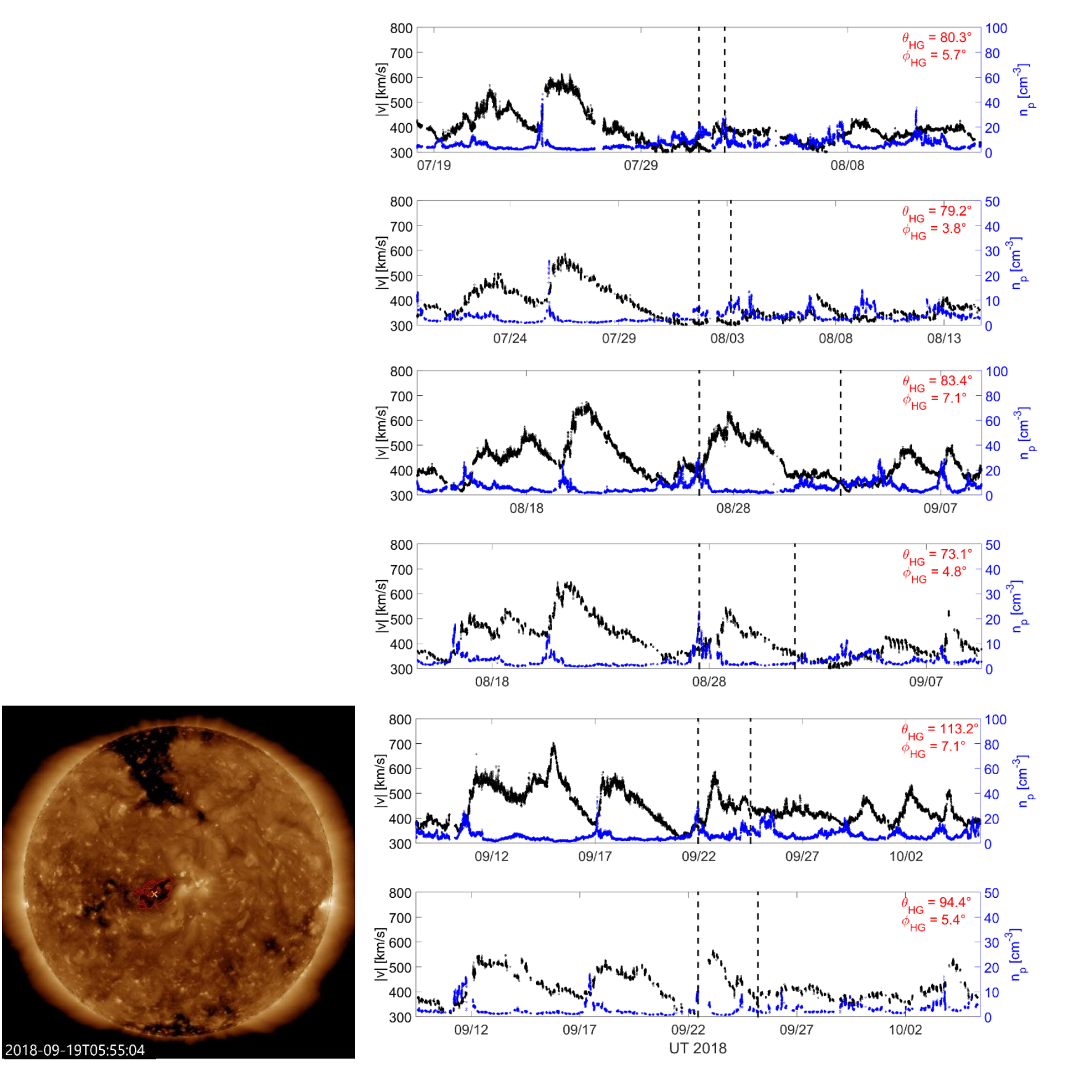}
\caption{Same as Fig.~\ref{1_2016} for CIR 7, occurring in 2018. No CH extraction was possible for the first two recurrences of the stream.}
\label{7_2018}
\end{figure*}

\begin{figure*}
\centering
\includegraphics[width = \textwidth]{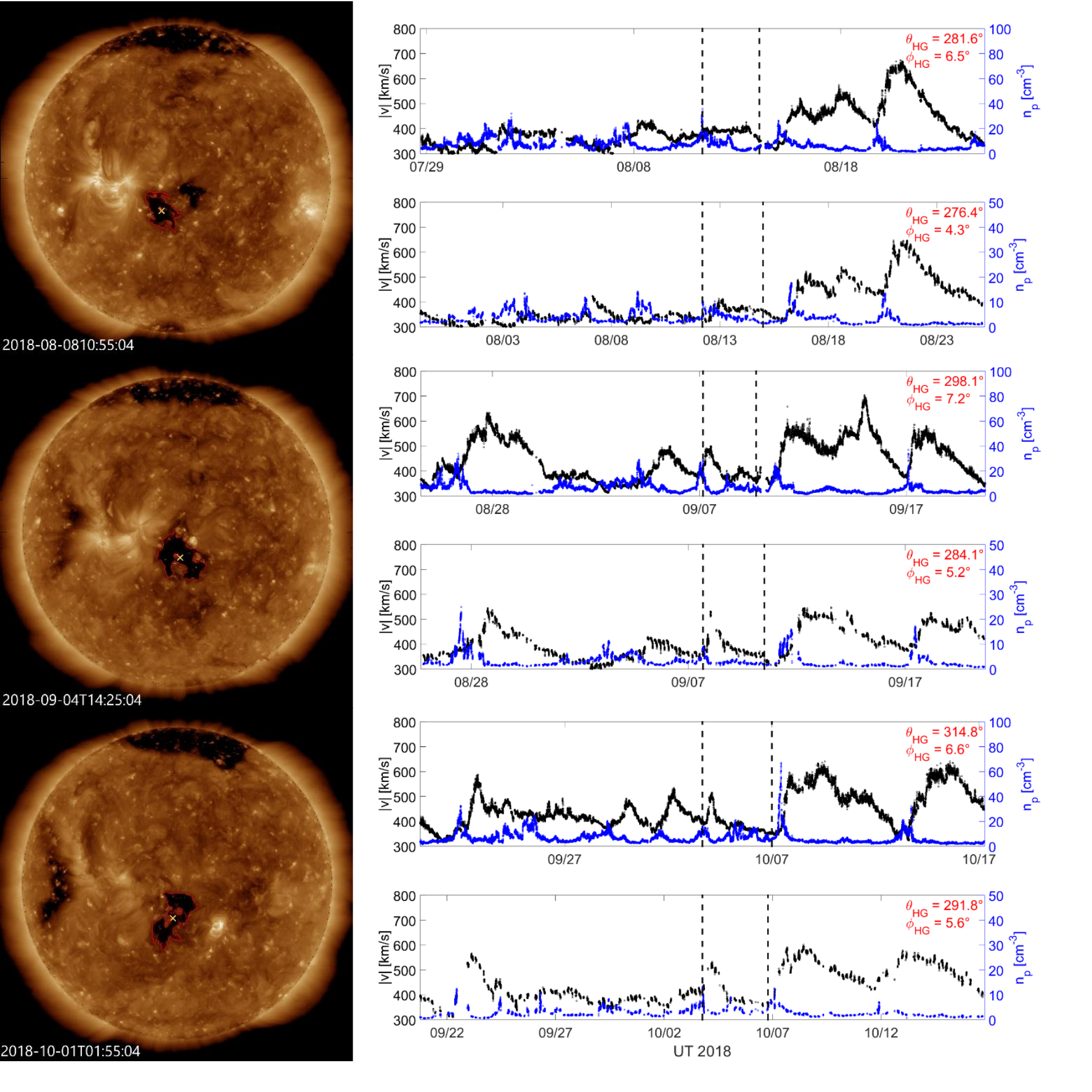}
\caption{Same as Fig.~\ref{1_2016} for CIR 8, occurring in 2018.}
\label{8_2018}
\end{figure*}

\end{appendix}

\end{document}